\documentclass[screen]{acmart}

\usepackage{ragged2e}
\usepackage{multirow}
\usepackage{multicol}
\usepackage{amsmath}
\usepackage{mathtools}
\definecolor{ao}{rgb}{0.0, 0.0, 0.8}
\definecolor{cyan4}{RGB}{0,139,139}
\definecolor{navy}{RGB}{0,0,128}
\usepackage[most]{tcolorbox}
\usepackage{tablefootnote}
\usepackage{booktabs}
\usepackage{array}
\usepackage{colortbl}
\usepackage{color}
\usepackage{xcolor}
\usepackage{listings}
\usepackage{hyperref}
\usepackage{fontawesome5}

\newcolumntype{L}[1]{>{\raggedright\let\newline\\\arraybackslash\hspace{0pt}}m{#1}}
\newcolumntype{C}[1]{>{\centering\let\newline\\\arraybackslash\hspace{0pt}}m{#1}}
\newcolumntype{R}[1]{>{\raggedleft\let\newline\\\arraybackslash\hspace{0pt}}m{#1}}

\newtcbox{\catbox}[1][red]{on line,
    colback=#1, colframe=#1, boxsep=0pt, boxrule=0pt, size=small, arc=0.7mm, left=-1pt, right=-1pt, top=-1pt, bottom=-0.5pt}

\newcommand{\setup}{\catbox[blue!10]{\faCog~\textsc{Setup} \texttt{\&} \textsc{Initialization}}}
\newcommand{\build}{\catbox[orange!12]{\faToolbox~\textsc{Building} \texttt{\&} \textsc{Packaging}}}
\newcommand{\test}{\catbox[green!10]{\faCheckDouble~\textsc{Test} \texttt{\&} \textsc{Quality Checks}}}
\newcommand{\deploy}{\catbox[red!10]{\faArrowAltCircleUp~\textsc{Deployment} \texttt{\&} \textsc{Post-Actions}}}
\newcommand{\report}{\catbox[gray!12]{\faEnvelope~\textsc{Notifications} \texttt{\&} \textsc{Reporting}}}

\newtcbtheorem{Summary}{\bfseries Summary of RQ}{enhanced,
	coltitle=black,
	top=0.1in,
	attach boxed title to top left=
	{xshift=1.5em,yshift=-\tcboxedtitleheight/2},
	boxed title style={size=small,colback=lightgray}
}{summary}

\newtcbtheorem{Summary_data_charac}{Summary}{enhanced,
	coltitle=black,
	top=0.1in,
	attach boxed title to top left=
	{xshift=1.5em,yshift=-\tcboxedtitleheight/2},
	boxed title style={size=small,colback=gray}
}{Summary_data_charac}

\newtcolorbox[auto counter]{summary}[1][]{title={\bfseries Summary},enhanced,
	coltitle=black,
	top=0.17in,
	attach boxed title to top left=
	{xshift=1.5em,yshift=-\tcboxedtitleheight/2},
	boxed title style={size=small,colback=lightgray},#1}

\newtcolorbox[auto counter]{summary_RQ1}[1][]{title={\bfseries Summary of RQ1},enhanced,
	coltitle=black,
	top=0.17in,
	attach boxed title to top left=
	{xshift=1.5em,yshift=-\tcboxedtitleheight/2},
	boxed title style={size=small,colback=lightgray},#1}

\newtcolorbox[auto counter]{summary_RQ2}[1][]{title={\bfseries Summary of RQ2},enhanced,
	coltitle=black,
	top=0.17in,
	attach boxed title to top left=
	{xshift=1.5em,yshift=-\tcboxedtitleheight/2},
	boxed title style={size=small,colback=lightgray},#1}

\newtcolorbox[auto counter]{summary_RQ3}[1][]{title={\bfseries Summary of RQ3},enhanced,
	coltitle=black,
	top=0.17in,
	attach boxed title to top left=
	{xshift=1.5em,yshift=-\tcboxedtitleheight/2},
	boxed title style={size=small,colback=lightgray},#1}
	
\newcommand{\yml}{\texttt{YML} }
\definecolor{codegreen}{rgb}{0,0.6,0}
\definecolor{codegray}{rgb}{0.5,0.5,0.5}

\lstdefinestyle{yml}{
     float=tp,
     floatplacement=tbp,
     abovecaptionskip=-5pt,
     numberstyle=\normalfont\tiny\color{gray},
     basicstyle=\color{black}\footnotesize\ttfamily,
     rulecolor=\color{black},
     string=[s]{'}{'},
     keywordstyle=\color{red},
     morecomment=[l]{-},
     morecomment=[l]{+},
     moredelim=[is][\color{red}]{|<}{>|},
     moredelim=[is][\color{blue}]{|>}{<|},
}

\newcommand{\up}{$\nearrow$}
\newcommand{\dn}{$\searrow$}

\usepackage{marginnote}
\renewcommand{\marginpar}{\marginnote}
\usepackage[backgroundcolor=white,bordercolor=blue,linecolor=blue]{todonotes}
\AtBeginDocument{	\providecommand\BibTeX{{			\normalfont B\kern-0.5em{\scshape i\kern-0.25em b}\kern-0.8em\TeX}}}
\begin{document}

\newcommand{\ourtitle}{CI/CD{\large~}Configuration{\large~}Practices{\large~}in{\normalsize~}Open-Source Android{\normalsize~}Apps:{\normalsize~}An{\large~}Empirical{\Large~}Study}

\title{\ourtitle}

\author{Taher A. Ghaleb}
\affiliation{
	\department{Department of Computer Science}
	\institution{Trent University}
	\city{Peterborough, ON}
	\country{Canada}
         \streetaddress{ (part of this work was done at the Faculty of Information, University of Toronto, Toronto, ON, Canada)}}
        \email{taherghaleb@trentu.ca}
    
\author{Osamah Abduljalil}
\affiliation{
	\department{Computer Science Department}
	\institution{Imam Mohammad Ibn Saud Islamic University}
	\city{Riyadh}
	\country{Saudi Arabia}}
        \email{osamah.abduljalil@gmail.com}
    
\author{Safwat Hassan}
\affiliation{
	\department{Faculty of Information}
	\institution{University of Toronto}
	\city{Toronto, ON}
	\country{Canada}}
        \email{safwat.hassan@utoronto.ca}
    
\renewcommand{\shortauthors}{Ghaleb et al.}

\begin{abstract}
Continuous Integration and Continuous Delivery (CI/CD) is a well-established practice that automatically builds, tests, packages, and deploys software systems.
To adopt CI/CD, software developers need to configure their projects using dedicated \yml configuration files.
Mobile apps have distinct characteristics with respect to CI/CD practices, such as testing on various emulators and deploying to app stores.
However, little is known about the challenges and added value of adopting CI/CD in mobile apps and how developers maintain such a practice.
In this paper, we conduct an empirical study on CI/CD practices in $2,557$ Android apps adopting four popular CI/CD services, namely \textsf{GitHub~Actions}, \textsf{Travis~CI}, \textsf{CircleCI}, and \textsf{GitLab~CI/CD}.
We also compare our findings with those reported in prior research on general CI/CD practices to situate them within broader trends.
We observe a lack of commonality and standardization across CI/CD services and Android apps, leading to complex \yml configurations and associated maintenance efforts. We also observe that CI/CD configurations focus primarily on the build setup, with around half of the projects performing standard testing and only $9\%$ incorporating deployment. In addition, we find that CI/CD configurations are changed bi-monthly on average, with frequent maintenance correlating with active issue tracking, project size/age, and community engagement.
Our qualitative analysis of commits uncovered $11$ themes in CI/CD maintenance activities, with over a third of the changes focusing on improving workflows and fixing build issues, whereas another third involves updating the build environment, tools, and dependencies.
Our study emphasizes the necessity for automation and AI-powered tools to improve CI/CD processes for mobile apps and advocates creating adaptable open-source tools to efficiently manage resources, especially in testing and deployment.
\end{abstract}



\maketitle

\section{Introduction}\label{sec:intro}
Continuous Integration and Continuous Delivery (CI/CD) is a well-established practice that automates the building, testing, packaging, and deployment of software systems~\cite{fowler2006continuous,Difference_Between_CI_CD,Difference_Between_CI_CD2}. CI/CD pipelines consist of automated processes designed to reduce human error, improve developer productivity, and accelerate release cycles. Their widespread adoption in both industry and open-source communities has led to extensive research on the challenges and benefits associated with CI/CD pipelines~\cite{hilton2017trade,hilton2016usage,thatikonda2023beyond}.

Mobile applications, henceforth referred to as \textit{mobile apps}, have become integral to our daily lives, serving various purposes, from communication to entertainment. Research reports that apps with more frequent release cycles tend to achieve higher ratings, as users often positively perceive regular updates and improvements~\cite{DBLP:journals/ese/McIlroyAH16}. Research has also explored the benefits of adopting CI/CD in the software development lifecycle, notably enhancing quality and accelerating the delivery timeline. In the context of CI/CD, \citet{DBLP:journals/ese/CruzAL19} studied the testing practices in open-source mobile apps and found that the use of CI/CD pipelines is surprisingly limited among the mobile apps they analyzed.
Mobile apps differ substantially from general-purpose software systems (e.g., desktop applications or web services) due to their unique ecosystem~\cite{DBLP:journals/ese/CruzAL19,hu2011automating,picco2014software}. Developers must handle complex user interactions (e.g., swipe, pinch)~\cite{zaeem2014automated} and account for devices with resource constraints, such as battery and lower processing power. Moreover, mobile apps need to support a wide range of operating system versions and hardware configurations~\cite{DBLP:conf/sigsoft/KhalidNSH14}, which often causes traditional testing tools to fall short in this context~\cite{wang2015mobile,maji2010characterizing}. Given that mobile apps typically follow a tight weekly or bi-weekly release cycle~\cite{DBLP:conf/wcre/NayebiAR16}, this can create pressure on testing and deployment tasks.
In the context of CI/CD pipelines, while regular applications might run tests and deploy to cloud servers, mobile apps require additional steps that demand platform-specific tooling and plugins, such as multi-variant builds, app bundle generation (e.g., \texttt{bundleRelease}), emulator-based instrumentation testing (e.g., \texttt{reactivecircus/android-emulator-runner}), code signing (e.g., \texttt{signingConfigs}), and deployment to mobile platforms like the \textsf{Google Play Store} (e.g., \texttt{actions/upload-artifact} or \texttt{fastlane}). These tasks introduce distinct configuration challenges, which our study aims to investigate.

Despite such importance, challenges in configuring and maintaining CI/CD for mobile apps have not been well studied. Though \citet{DBLP:conf/kbse/LiuSZL0022} studied the adoption of CI/CD in mobile apps, their study mainly provided general statistical insights about CI/CD usage without exploring in-depth the specific configuration or maintenance practices of developers.
In this paper, we conduct an empirical study to investigate the configuration practices and maintenance activities associated with CI/CD in a curated set of 2,557 open-source Android apps adopting four popular CI/CD services, namely \textsf{GitHub~Actions}, \textsf{Travis~CI}, \textsf{CircleCI}, and \textsf{GitLab~CI/CD}. In particular, our study explores the common patterns and challenges that developers face with CI/CD configurations for their apps. To contextualize our observations within the broader trends of CI/CD, we compare and contrast our findings with previous research findings on general-purpose CI/CD practices. We aim to provide insights into the current state of CI/CD adoption within the Android development community and highlight areas for potential improvement.
We address the following research questions (RQs).

\vspace{4pt}
\noindent\textbf{RQ1: How do CI/CD configurations vary in complexity and tool usage for Android apps?}

\noindent In this RQ, we examine CI/CD configurations across multiple services. We observe that CI/CD configurations are not standardized, with complexities varying by service and type of configuration. We also find that most of the attention in CI/CD configuration is on the pipeline setup, accounting for more than 81\% of the configurations. While testing is found to be present in about half of the projects, mainly through standard unit tests, deployment is comparatively less common, occurring in 9\% of the projects. Our investigation shows that configuring deployment is notably complex and often does not directly target app stores, indicating a need for more integrated and automated solutions that streamline mobile app deployments to popular platforms, such as the \textsf{Google Play Store}.

\vspace{4pt}
\noindent\textbf{RQ2: How do CI/CD configurations evolve in Android apps?}

\noindent Adopting CI/CD in mobile app development makes developers responsible for the ongoing maintenance of their pipelines, which involves tasks such as feature additions and issue resolutions.
In this RQ, we analyze and model the evolution of CI/CD configurations.
We observe that developers regularly maintain CI/CD pipelines, yet not very frequently, mainly focusing on improving the setup and build phases. Over time, projects tend to simplify their CI/CD pipelines. While maintenance is found to occur bi-monthly on average, many projects update CI/CD pipelines more often, especially those using \textsf{GitHub~Actions}. Our results show that increased maintenance activity correlates significantly with factors such as active issue tracking, project size/age, and community engagement.

\vspace{4pt}
\noindent\textbf{RQ3: What are the common themes in CI/CD maintenance activities in Android apps?}

\noindent As CI/CD pipelines are continually updated, it is important to understand the rationale for these changes. In this RQ, we perform a manual analysis of a statistically significant random sample of commits related to CI/CD configurations. We observe that more than a third of CI/CD configuration changes focus on improving workflow and fixing build issues, while another third targets updates to the build environment, tools, and dependencies. To validate our manually identified themes for all commits in our dataset, we employ Latent Dirichlet Allocation (LDA) for automated topic generation and observe that the results are consistent with our observations.

\noindent The key contributions of this paper are as follows.
\begin{itemize}
    \vspace{-8pt}
    \item This study is the first of its kind in analyzing CI/CD configurations in Android apps across various services, using a dataset of 2,557 projects. It highlights current challenges in CI/CD adoption due to non-standardized practices.

    \item This study models the evolution of CI/CD configurations using linear regression and identifies the potentially relevant factors related to changes in CI/CD pipelines.

    \item This study presents a manually validated classification of 11 themes related to the maintenance of CI/CD pipelines, corroborated by automatically generated topics using LDA, ensuring both depth and breadth in our analysis.
    
    \item This study contextualizes the CI/CD trends and practices observed in Android apps with prior findings on general-purpose CI/CD practices in open-source and industrial applications.

\end{itemize}
 
\noindent Overall, our study highlights the lack of standardized configurations in CI/CD pipelines for Android apps, emphasizing the need for uniform practices to improve interoperability. In addition, this study recommends employing automation tools and AI-driven solutions, such as bots and intelligent recommendation systems, to enhance and streamline CI/CD processes. Moreover, researchers are encouraged to develop adaptable frameworks and open-source tools to address resource constraints in mobile apps, especially those concerned with testing and deployment phases.

\vspace{4pt}
\noindent\textbf{Paper organization:} The rest of this paper is organized as follows. Section~\ref{sec:bg} provides background information relevant to our study.
Section~\ref{sec:Study_Setup} details our study setup to investigate the challenges in configuring CI/CD pipelines for mobile apps.
Section~\ref{sec:Results} outlines our research questions, providing both their motivation and the obtained results.
We discuss the implications of our work in Section~\ref{sec:Implications}. 
Section~\ref{sec:Threats} illustrates the potential validity threats of our findings.
We present the related work in Section~\ref{sec:RelatedWork}. 
Finally, we conclude our work and provide recommendations for future work in Section~\ref{sec:Conclusion}.

\section{Background}\label{sec:bg}
Continuous Integration (CI) and Continuous Delivery (CD) are essential practices in modern software development, particularly for Android mobile apps~\cite{tak2018mobile}. These practices ensure that code changes are automatically built, tested, and prepared for release, allowing for faster and more reliable app updates.

Continuous integration allows developers to frequently commit code changes to a shared, central repository, triggering automated build and test processes for smooth integration~\cite{fowler2006continuous}. This helps catch bugs early, streamline new features, and improve code quality. 
Continuous delivery extends CI by automating deployment and preparing code changes for release~\cite{humble2010continuous}. The deployment process in CI/CD involves moving integrated and tested changes to a production environment. Deployments can be manual or automated, depending on project needs. Automated deployments reduce the risk of human error and ensure consistency across different environments.
Many CI/CD service providers (referred to as CI/CD services) are available to streamline the development process for Android apps. Examples include: \textsf{GitHub~Actions}\footnote{\url{https://docs.github.com/en/actions}}
, \textsf{Travis~CI}\footnote{\url{https://docs.travis-ci.com}}, \textsf{CircleCI}\footnote{\url{https://circleci.com/docs}}, and \textsf{GitLab~CI/CD}\footnote{\url{https://docs.gitlab.com/ci}}. Each service offers unique features and configurations tailored to various project needs and development workflows.
Triggering CI/CD events can be done in several ways. Some services, like \textsf{GitHub~Actions}, have native integrations that allow seamless triggering of CI/CD events within the same platform. Other services, like \textsf{CircleCI}, enable CI/CD builds to be triggered on third-party servers using \textit{webhooks} for specific events, such as when code is pushed or a pull request is created.
In this section, we describe the CI/CD process in general, how it is configured in the context of Android mobile app development, and the differences in CI/CD configuration across various services.

\subsection{CI/CD Process}
\begin{figure*}
	\includegraphics[width=\textwidth]{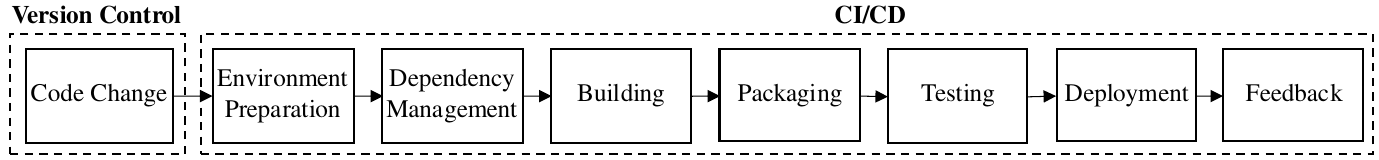}
        \vspace{-15pt}
	\caption{General overview of the CI/CD process.}
        \vspace{-10pt}
	\label{fig:ci-cd-process}
\end{figure*}

In an Android CI/CD pipeline, various processes are involved to ensure seamless integration and delivery. Figure~\ref{fig:ci-cd-process} gives an overview of the CI/CD process in general. These processes are typically configured in \yml files and include: \textit{Environment Setup}, \textit{Dependency Management}, \textit{Compiling/Building}, \textit{Packaging}, \textit{Testing}, \textit{Deployment}, and \textit{Feedback}.
In Listing~\ref{lst:yml}, we provide a sample \yml configuration for a CI/CD pipeline (in \textsf{GitHub~Actions}) and give a brief description of each process. The {\color{red}\textbf{\texttt{red}}} text represents the CI/CD \textit{directives} (keys that define specific instructions) that guide the build, {\color{blue}\textbf{\texttt{blue}}} text represents executable commands, {\color{black}\textbf{\texttt{black}}} text represents parameter values, and {\color{codegreen}\textbf{\texttt{green}}} text represents comments.

\begin{itemize}
    \item \textbf{Environment Preparation.} In this stage, the CI/CD service initializes the environment needed to build and test an Android app, which involves choosing the machine operating system, specifying the appropriate version of the Java Development Kit (JDK), setting up Android SDK\footnote{\url{https://developer.android.com/studio}} components, and configuring any necessary environment variables (see \textit{lines 5-12}).
    
    \item \textbf{Dependency Management.} In this stage, the dependencies required to build and run an Android app are installed, which typically involves running tasks for the \textit{Gradle}\footnote{\url{https://gradle.org}} build management tool, for example, to download and cache dependencies specified in build files (see \textit{lines 14-15}).
    
    \item \textbf{Building.} In this stage, an Android app is compiled by running respective build tasks to compile the source code and generate executables and resources (see \textit{lines 17-18}).

    \item \textbf{Packaging.} In this stage, an Android app is packaged into a certain distribution format, such as the Android Package (APK)\footnote{\url{https://developer.android.com/studio/profile/apk-profiler}} or the Android App Bundle (AAB)\footnote{\url{https://developer.android.com/guide/app-bundle}}, which comprises executables with any required configurations and third-party libraries (see \textit{lines 20-25}).

    \item \textbf{Testing.} In this stage, automated tests are executed to ensure that the application works as expected, which can include unit tests, integration tests, and user interface tests whose results determine whether the build can proceed to the next stage (see \textit{lines 27-28}).
    
    \item \textbf{Deployment.} In this stage, the successfully built and tested application is deployed to a production environment, which typically involves uploading the APK file to a distribution platform, such as the \textsf{Google Play Store}\footnote{\url{https://play.google.com}} (see \textit{lines 30-33}).

    \item \textbf{Feedback.} After all, feedback on the build and release process is sent back to developers, which involves build logs, test results, and any issues or areas for improvement (see \textit{lines 35-39}).

\end{itemize}

\begin{lstlisting}[style=yml,caption={Sample YML configuration file (GitHub~Actions)}, label={lst:yml},numbers=left,frame=lines,escapechar=\!]
|<name>|: Android CI
|<on>|: [push, pull_request]
|<jobs>|:
  |<build>|:
    |<runs-on>|: ubuntu-latest                !{\color{codegreen}\# (1) Environment Setup}!
    |<steps>|:
    |<- name>|: Checkout code                   
      |<uses>|: actions/checkout@v2             
    |<- name>|: Set up JDK and SDK              
      |<run>|: |
        |>actions/setup-java@v2<| --java-version '11'
        |>sdkmanager<| "platforms;android-30" "build-tools;30.0.3"

    |<- name>|: Install dependencies          !{\color{codegreen}\# (2) Dependency Management}!
      |<run>|: |>./gradlew<| dependencies

    |<- name>|: Build                         !{\color{codegreen}\# (3) Building}!
      |<run>|: |>./gradlew<| build assembleDebug

    |<- name>|: Generate APK                  !{\color{codegreen}\# (4) Packaging}!
      |<if>|: success()
      |<uses>|: actions/upload-artifact@v3
      |<with>|:
        |<name>|: app.apk
        |<path>|: app/build/apk/app.apk

    |<- name>|: Test                          !{\color{codegreen}\# (5) Testing}!
      |<run>|: |>./gradlew<| test

    |<- name>|: Deploy to Google Play         !{\color{codegreen}\# (6) Deployment}!
      |<run>|: |
        |>echo<| $ANDROID_KEYSTORE | base64 --decode > keystore.jks
        |>./gradlew<| publish --no-daemon

    |<- name>|: Collect Logs and Notify       !{\color{codegreen}\# (7) Feedback}!
      |<run>|: |
        |>./gradlew<| collectLogs
        |>curl<| -X POST -H 'Content-type: application/json' --data
                 '{"text":"Completed successfully."}' $SLACK_WEBHOOK_URL
\end{lstlisting}

\subsection{CI/CD Configuration in Different Services}
Configuring CI/CD pipelines varies across different services. Here, we highlight the main differences to explain how these variations might influence the overall developer's perception and experience with each of the four CI/CD services.
\subsubsection{GitHub~Actions} A CI/CD service that is directly integrated within \textsf{GitHub}, allowing developers to define their CI/CD workflows using \yml files placed in the \texttt{.github/workflows} directory. This service is highly flexible, enabling developers to customize every step of the CI/CD process using \textit{job} and \textit{step} directives. Its integration with \textsf{GitHub} repositories facilitates the triggering of workflows based on various events (e.g., pushes, pull requests, and more). In addition, \textsf{GitHub~Actions} provides access to a wide range of pre-built actions available in the \textsf{GitHub Marketplace}, which can be easily incorporated into workflows.

\subsubsection{Travis~CI}
An independent CI/CD service that simplifies the CI/CD setup with its use of a \texttt{.travis.yml} file located in the root of the repository. It employs predefined directives, such as \texttt{install}, \texttt{script}, \texttt{before\_script}, and \texttt{after\_success}, making it straightforward to set up and use. This allows developers to quickly establish a CI/CD pipeline without needing to define every process explicitly. \textsf{Travis~CI} integrates seamlessly with \textsf{GitHub} repositories, making it a popular choice for many open-source projects due to its simplicity and ease of use.

\subsubsection{CircleCI}
An independent CI/CD service that uses a \texttt{circle.yml} or \texttt{config.yml} file found in the \texttt{.circleci/} directory. It provides predefined jobs like \texttt{checkout}, \texttt{run}, and \texttt{deploy}, which help structure the CI/CD process while allowing for the inclusion of custom scripts for more complex setups. \textsf{CircleCI} is known for its strong support for \textsf{Docker}, enabling containerized builds that ensure consistency across different environments\footnote{\url{https://circleci.com/docs/about-circleci}}. This makes \textsf{CircleCI} suitable for projects requiring advanced build and deployment configurations, such as Android mobile apps.

\subsubsection{GitLab~CI/CD}
A fully integrated within the \textsf{GitLab} version control system, but can also be used within \textsf{GitHub}. It uses a \texttt{.gitlab-ci.yml} file to define the CI/CD pipeline. Similar to \textsf{Travis~CI}, it features predefined processes such as \texttt{before\_script}, \texttt{script}, and \texttt{after\_script}, while also allowing developers to create custom processes and jobs for greater flexibility. The integration with \textsf{GitHub} repositories is straightforward, allowing developers to connect their \textsf{GitHub} projects to \textsf{GitLab~CI/CD} for continuous integration and deployment. This capability makes it a versatile option for teams that use \textsf{GitHub} for source control but prefer \textsf{GitLab}'s CI/CD features. Moreover, \textsf{GitLab~CI/CD} supports extensive customization, making it ideal for projects that require intricate workflows and detailed stage definitions.

\section{Study setup}\label{sec:Study_Setup}
This section presents our study setup. 
We explain how we collect and prepare the data for our studied RQs. Our data, scripts, and raw results can be found in our replication package~\cite{our_replication_package}.

\subsection{Data Collection}
Figure~\ref{fig:data_collection} provides an overview of our data collection process. We focus on Android apps given the extensive research on them in the literature and the availability of public and open-source data. 
For data collection, we rely on two primary sources that have been commonly used in prior research: \textsf{GitHub}\footnote{https://github.com} and the \textsf{F-Droid}\footnote{https://www.f-droid.org} store~\cite{cruz2019attention,lin2020test,poll2022automating,wang2023empirical}.

\subsubsection{Collecting Android projects from \textsf{F-Droid}}
The \textsf{F-Droid}~\cite{FDroid} store serves as a prominent repository for Android apps, offering a diverse selection of mobile apps. 
\textsf{GitHub} projects were previously studied in prior work as \textsf{GitHub} offers extensive data for analyzing developer activities, such as the commit messages and reported issues.
Hence, we select apps that contain links to their \textsf{GitHub} project. Following this filtration, we identify $3,911$ mobile apps.

\subsubsection{Collecting Android projects from \textsf{GitHub}}
To gather data from \textsf{GitHub}, we utilize the official \textsf{GitHub} API\footnote{https://docs.github.com/en/rest}. 
Given the enormous base of \textsf{GitHub} projects, we focus specifically on projects that are classified under the \texttt{`android'} topic\footnote{https://github.com/topics/android}, similar to \citet{liu2020androzooopen}. 
This step, conducted in May 2023, provides us with $108k$ \textsf{GitHub} projects.
To further refine our dataset, we exclude projects without a user interface by verifying the presence of an \texttt{AndroidManifest.xml} file with an \texttt{activity} tag. 
This filtration step ensures that the selected projects are indeed Android apps, thereby enhancing the accuracy and relevance of our dataset for analysis. At the end of this step, we identify $40,886$ Android-related projects. 

\begin{figure*}
	\includegraphics[width=\textwidth]{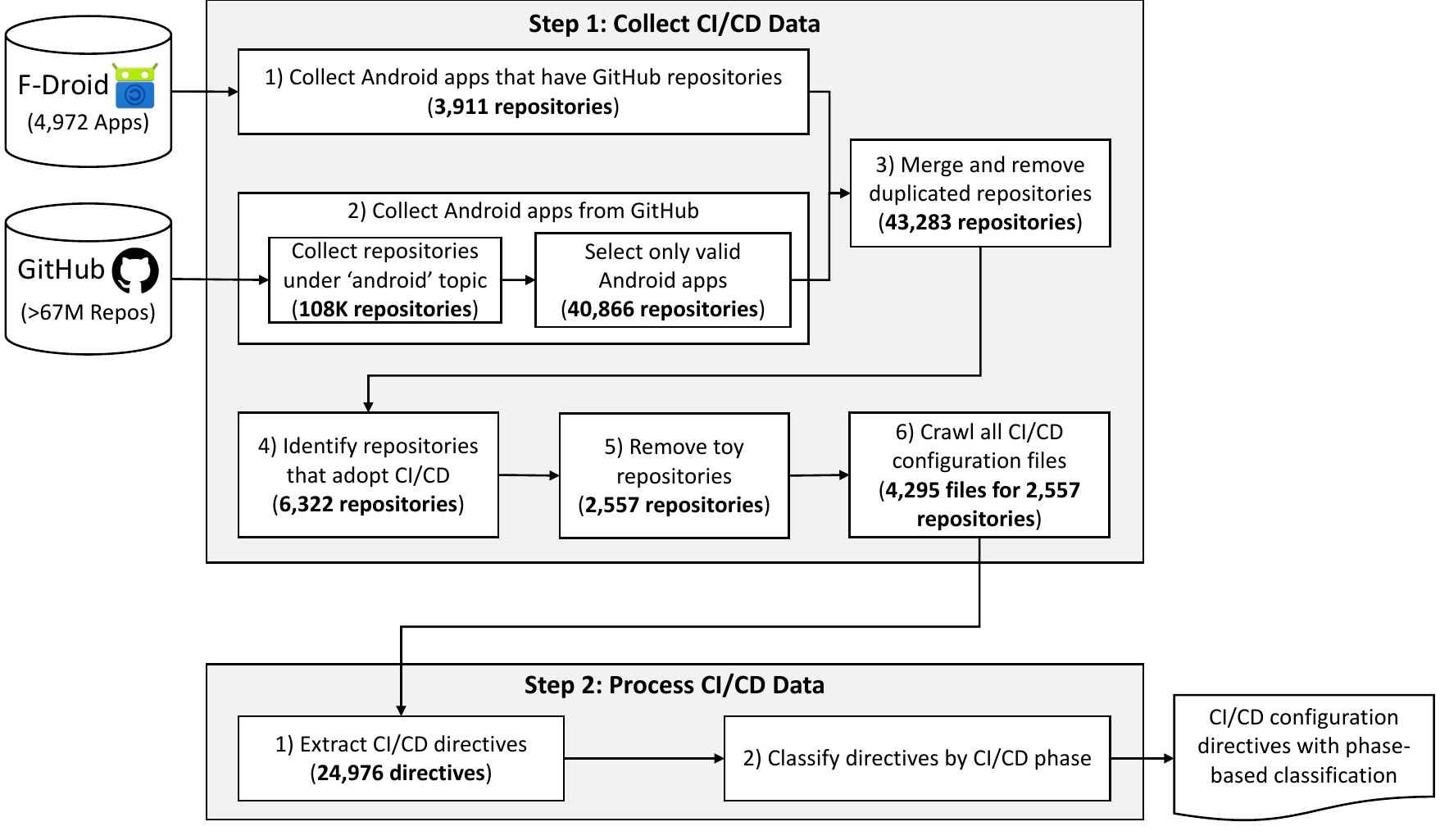}
	\caption{An overview of our data collection process}
	\label{fig:data_collection}  
\end{figure*}

\subsubsection{Merging all projects from the two sources}
After collecting Android projects from both \textsf{GitHub} and the \textsf{F-Droid} store, we remove duplicate projects and those that have zero \textsf{GitHub} stars, and identify $43,283$ unique projects.

\subsubsection{Identifying projects that adopt CI/CD} 
CI/CD pipelines are commonly configured using \yml files (i.e., files ending with \texttt{`.yml'} or \texttt{`.yaml'}), which are used to define instructions for automating the build, test, and deployment steps.
We explore the available documentation of popular CI/CD services (e.g., \textsf{GitHub~Actions}) to identify the name and location of the needed \yml files to define the CI/CD pipeline in each service.
Table~\ref{tab:yml_file_path} presents a list of the identified \yml files alongside the respective CI/CD services to which they belong. 
Finally, using the \textsf{GitHub API}, we identify projects containing \yml files. We identify $6,322$ projects that adopt CI/CD practices. 

\begin{table}[ht]
\centering
\caption{List of CI/CD services with their corresponding \yml file path}
\label{tab:yml_file_path}
\begin{tabular}{ l  l }
\toprule
CI/CD service & Path (regular expression) \\
\midrule
\textsf{GitHub~Actions} & \texttt{r'\textbackslash{}.github/workflows/.*\textbackslash{}.(yml|yaml)\$'} \\

\textsf{Travis~CI} & \texttt{r'.travis\textbackslash{}.yml\$'} \\

\textsf{CircleCI} & \texttt{r'(\textbackslash{}.circleci/config\textbackslash{}.yml|circle\textbackslash{}.yml)\$'} \\

\textsf{GitLab~CI/CD} & \texttt{r'.gitlab-ci\textbackslash{}.yml\$'} \\

\textsf{Azure Pipelines} & \texttt{r'azure-pipelines\textbackslash{}.yml\$'} \\

\textsf{AppVeyor} & \texttt{r'(.appveyor\.yml|appveyor\.yml)\$'} \\
\textsf{Bitbucket} & \texttt{r'bitbucket-pipelines\textbackslash{}.yml\$'} \\
\bottomrule
\end{tabular}
\end{table}

\subsubsection{Removing toy projects}
Android projects containing \yml files do not necessarily represent an actual mobile app. 
For example, tutorial and library projects include example apps developed for demonstration purposes that have a structure identical to actual mobile apps, including \texttt{AndroidManifest.xml} and \yml files, but they are not maintained or updated further.
We perform manual and automated analysis to identify and exclude toy projects from our dataset. 
First, through collaborative discussion sessions, two raters
(co-authors of this paper with practical experience in software engineering and Android app development)
manually examined a statistically significant random sample of 93 projects (a confidence level of $95\%$ and a confidence interval of $\pm 10$), reviewing their repository descriptions and \texttt{AndroidManifest.xml} paths to identify an initial set of indicative keywords (e.g., \textit{`example'}, \textit{`sample'}, \textit{`library'}, and \textit{`tutorial'}). 
Based on this analysis, we develop heuristic rules in the form of a Python script to automate the detection of toy projects within our dataset using these keywords.

To assess the effectiveness of our automated filtering, the same two co-authors independently validate another statistically significant random sample of 93 projects. Each rater indicates whether there is a match or mismatch between the automated label and their manual judgment. Disagreements or mismatches were discussed in consensus meetings, with input from the third co-author when needed. In these two initial iterations, we observed $14\%$-$19\%$ mismatch rates between the manual and automated labeling, prompting refinement of our heuristics. This process of validation, refinement, and re-evaluation was repeated iteratively six times for the remaining unfiltered projects until a $93\%$ accuracy rate was achieved in manually validating the automatically filtered projects, ensuring that our final dataset reliably excluded toy projects and retained real Android mobile apps.
After removing toy projects, we identify $2,719$ projects that adopt CI/CD practices. 

We exclude CI/CD services (e.g., \textsf{AppVeyor}, \textsf{Azure Pipelines}, and \textsf{Bitbucket}) with fewer than $1.00\%$ projects, as this percentage is not representative enough to draw conclusions.
Finally, we obtain $2,557$ Android apps adopting four different CI/CD services, namely \textsf{GitHub~Actions}\footnote{\url{https://github.com/features/actions}}, \textsf{Travis~CI}\footnote{\url{https://www.travis-ci.com}}, \textsf{CircleCI}\footnote{\url{https://circleci.com}}, and \textsf{GitLab~CI/CD}\footnote{\url{https://docs.gitlab.com/ee/ci}}.  
The majority (about $90\%$) of these projects are developed using the \textit{Kotlin}, \textit{Java}, and \textit{Dart} programming languages. Table~\ref{tab:langs} shows the full list of programming languages with their corresponding number of projects across CI/CD services.

\begin{figure}
    \centering
        \begin{minipage}{1\textwidth}
            \includegraphics[width=1\linewidth]{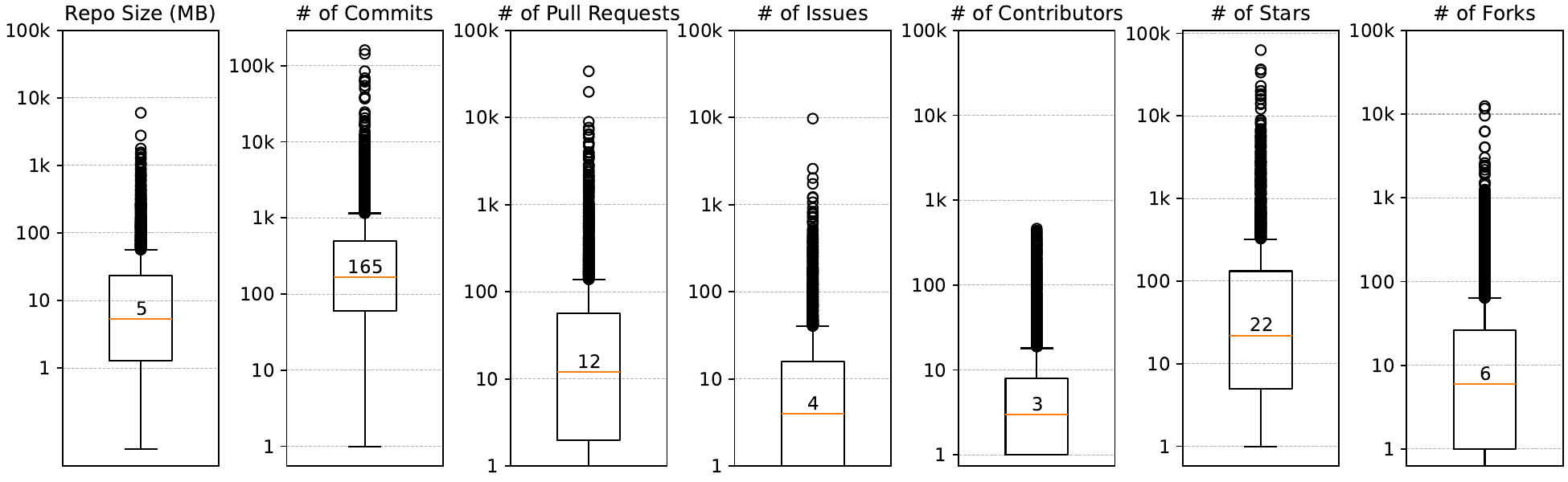}
            \vspace{-5pt}
            \caption{Boxplots showing log-scaled distributions of repository-level metrics the $2,557$ studied projects, with medians annotated and highlighted in orange color}
            \label{fig:bloxplots_projects_statistics}
        \end{minipage}
    \vspace{-10pt}
\end{figure}

Table~\ref{tab:projects_services} shows our final set of projects and CI/CD services. 
To provide an overview of our dataset, we present boxplots showing the distribution of key repository-level metrics for our studied projects in Figure~\ref{fig:bloxplots_projects_statistics}. Each boxplot summarizes one metric, such as the codebase size (in megabytes), number of commits, pull requests, issues, and popularity indicators like stars and forks. This visualization highlights the large variance across projects (with medians highlighted in orange), particularly in contribution and popularity metrics, with some repositories exhibiting orders-of-magnitude differences.

\subsubsection{Crawl all CI/CD configuration files}
For the final set of projects, we crawl the configuration \yml files associated with the adopted CI/CD services. We collect a total of $4,295$ files. We should note that (a) there are projects that adopt multiple services at the same time, and (b) other projects have multiple \yml files for one service. This led to having more configuration files compared to the number of projects (an average of 1.7 files per project).

\begin{table}[ht]
\centering
\caption{CI/CD services with their corresponding numbers of projects and \yml files}
\vspace{-5pt}
\label{tab:projects_services}
\begin{tabular}{p{4cm} r r}
\toprule
\textbf{CI/CD service} & \textbf{\# of projects} & \textbf{\# of \yml files} \\
\midrule
\textsf{GitHub~Actions} & 1,670 (65\%)  &  3,263 (76\%) \\
\textsf{Travis~CI}      & 755   (30\%)  &  748   (17\%) \\
\textsf{CircleCI}       & 198   (8\%)   &  196   (5\%) \\
\textsf{GitLab~CI/CD}   & 87     (3\%)  &  88    (2\%) \\
\midrule
\textsf{Total} & 2,557 & 4,295 \\
\bottomrule
\end{tabular}
\vspace{-1pt}
\end{table}

\begin{table}[ht]
\centering
\caption{Top 10 programming languages in our dataset with their corresponding number of projects across CI/CD services}
\vspace{-5pt}
\label{tab:langs}
\begin{tabular}{ l  r  r  r  r  r }
\toprule
\textbf{Language} & \textsf{\textbf{GitHub~Actions}} & \textsf{\textbf{Travis~CI}} & \textsf{\textbf{CircleCI}} & \textsf{\textbf{GitLab~CI/CD}} & \textbf{Total} \\
\midrule
Kotlin      & 770 & 188 & 109 & 24 & \textbf{1,091}\\
Java        & 477 & 461 & 62 & 44 & \textbf{1,044}\\
Dart        & 231 & 21 & 4 & 3 & \textbf{259}\\
C++         & 46 & 12 & 2 & 7 & \textbf{67}\\
JavaScript  & 36 & 18 & 4 & 1 & \textbf{59}\\
TypeScript  & 48 & 9 & 1 & 1 & \textbf{59}\\
C           & 21 & 16 & 1 & 4 & \textbf{42}\\
HTML        & 9 & 7 & 4 & 1 & \textbf{21}\\
C\#         & 6 & 1 & 1 & 0 & \textbf{8}\\
Objective-C & 3 & 2 & 2 & 1 & \textbf{8}\\
\bottomrule
\end{tabular}
\end{table}

\subsection{Data Processing}

\subsubsection{Extracting CI/CD configurations}
We extract all relevant \yml files representing CI/CD configurations from projects that adopt CI/CD practices. 
Subsequently, we process these files using the \textsc{PyYAML} package\footnote{\url{https://pypi.org/project/PyYAML}} in Python.
Each file contains pairs of keys (referred to as \textit{directives}) and values. Directives like `\texttt{name:}', `\texttt{run:}', and others are the keys that define specific instructions or configurations in a CI/CD workflow. Each directive has a corresponding value that specifies how the workflow should operate, such as the name of a job, a command to run, or other settings.
Through parsing each of these files, we extract the directives and their corresponding values by combining each directive with its preceding directives, thus representing their full hierarchical path within the configuration. For example, a value, such as ``\texttt{./gradlew dependencies}'', associated with a directive like `\texttt{jobs.build.steps.run}' instead of `\texttt{run}' alone, resulting in a complete path that distinguishes directives across the CI/CD process. This approach helps capture comprehensive information on the CI/CD setups implemented across various projects, making it easier to understand the context and connections of each directive within the overall configuration.
In total, our dataset contains a set of $24,976$ unique pairs of directives/values.

\subsubsection{Categorizing configuration directives}
\label{categorizing_directives}
To study the configuration characteristics among CI/CD pipelines, we manually classify the configuration directives into their respective phases or roles. To do this, we perform a combination of manual and automated analyses of directives in CI/CD configurations, similar to the way we use to filter out toy projects. Specifically, two raters
(co-authors of this paper with practical experience in software engineering and CI/CD practices)
perform a manual analysis of a statistically significant random sample of $96$ directives (a confidence level of $95\%$ and a confidence interval of $\pm 10$) through collaborative discussion sessions. Based on this initial analysis, we develop a Python script to automate the categorization process and apply it to the entire dataset. The automatically categorized directives are then manually validated by the same raters. This process is repeated iteratively until all directives are assigned to appropriate categories. Disagreements during validation are resolved through consensus meetings, in consultation with the third co-author.
We use the Cohen's $kappa$ inter-rater agreement statistic~\cite{cohen1960coefficient} to measure how reliable the manual validation of the automatically filtered projects is, and we achieve a strong inter-rater agreement ($k = 0.80$) and an observed agreement ($0.89$) in manually validating the automatically filtered projects.
This process results in five categories representing CI/CD directives: {\setup}, {\build}, {\test}, {\deploy}, and {\report}.

We should note that, while our categorization primarily relies on the functional phase to which each directive belongs, some directives are categorized based on their role within the configuration file, rather than the phase to which they belong. For example, within the above phases, some directives do not directly pertain to the functionality of a given phase (e.g., the \textit{id} or \textit{name} of that phase). Therefore, these types of directives are categorized under the {\setup} category.
In addition, for simplicity, we merge closely related phases in the CI/CD process. For example, directives related to \textit{Environment Setup} and \textit{Dependency Management} are both part of initialization and are thus categorized as {\setup}. Likewise, directives related to \textit{Building} and \textit{Packaging} are categorized as {\build}, as building source code is typically followed by packaging.

Below, we provide a description and examples for each category.

\begin{enumerate}
\item \textbf{\setup.} Configuration directives related to setting up the environment and initializing necessary resources. This includes installing dependencies, setting up environment variables, and preparing the build environment. Listing~\ref{lst:setup_example} gives an example of directives under this category in \textsf{GitHub~Actions} and \textsf{Travis~CI}.

\begin{table}[ht]
\caption*{Listing~\ref{lst:setup_example}: Example of directives under the {\setup} category}
\vspace{-6pt}
\begin{tabular}{C{7.2cm} | C{7.2cm}}
\multicolumn{1}{c}{\textsf{GitHub~Actions}} & \multicolumn{1}{c}{\textsf{Travis~CI}} \\
\hline
\captionsetup{labelformat=empty}
\lstset{style=yml}
\begin{lstlisting}[label={lst:setup_example},caption={~},aboveskip=-18pt,belowskip=-8pt]
|<steps>|:
  |<runs-on>|: ubuntu-latest
  |<- name>|: Set up JDK
    |<uses>|: actions/setup-java@v2
    |<with>|:
      |<java-version>|: '11'
\end{lstlisting}
&
\lstset{style=yml}
\begin{lstlisting}[aboveskip=-8pt,belowskip=-8pt]
|<language>|: android
|<jdk>|: openjdk11
|<android>|:
  |<components>|:
    - build-tools-30.0.3
    - android-30
\end{lstlisting}
\\\hline
\end{tabular}
\end{table}

\item \noindent\textbf{\build.} Configuration directives involved in the build process and packaging of an Android app. This typically includes compiling the source code and generating build artifacts. Listing~\ref{lst:build_example} gives an example of directives under this category in \textsf{GitHub~Actions} and \textsf{Travis~CI}.

\begin{table}[ht]
\centering
\caption*{Listing~\ref{lst:build_example}: Example of directives under the {\build} category}
\vspace{-6pt}
\begin{tabular}{C{7.2cm} | C{7.2cm}}
\multicolumn{1}{c}{\textsf{GitHub~Actions}} & \multicolumn{1}{c}{\textsf{Travis~CI}} \\
\hline
\captionsetup{labelformat=empty}
\lstset{style=yml}
\begin{lstlisting}[label={lst:build_example},caption={~},aboveskip=-8pt,belowskip=-8pt]
|<steps>|:
  |<- run>|: |>./gradlew<| build assembleDebug
\end{lstlisting}
&
\lstset{style=yml}
\begin{lstlisting}[aboveskip=-8pt,belowskip=-8pt]
|<script>|:
  |>- ./gradlew<| assembleDebug

\end{lstlisting}
\\\hline
\end{tabular}
\end{table}

\item \noindent\textbf{\test.} Configuration directives related to running tests and performing quality checks on the software. This includes executing unit tests, integration tests, and other quality assurance processes. Listing~\ref{lst:test_example} gives an example of directives under this category in \textsf{GitHub~Actions} and \textsf{Travis~CI}.

\begin{table}[ht]
\centering
\caption*{Listing~\ref{lst:test_example}: Example of directives under the {\test} category}
\vspace{-6pt}
\begin{tabular}{C{7.2cm} | C{7.2cm}}
\multicolumn{1}{c}{\textsf{GitHub~Actions}} & \multicolumn{1}{c}{\textsf{Travis~CI}} \\
\hline
\captionsetup{labelformat=empty}
\lstset{style=yml}
\begin{lstlisting}[label={lst:test_example},caption={~},aboveskip=-18pt,belowskip=-8pt]
|<steps>|:
  |<- run>|: |>./gradlew<| check test ktlint lint -S
\end{lstlisting}
&
\lstset{style=yml}
\begin{lstlisting}[aboveskip=-8pt,belowskip=-8pt]
|<script>|:
  |>- ./gradlew<| test

\end{lstlisting}
\\\hline
\end{tabular}
\end{table}

\item \noindent\textbf{\deploy.} Configuration directives for deploying the software to target environments and executing post-deployment actions. This includes steps to push the build artifacts to a production environment. Listing~\ref{lst:deploy_example} gives an example of directives under this category in \textsf{GitHub~Actions} and \textsf{Travis~CI}.

\begin{table}[ht]
\centering
\caption*{Listing~\ref{lst:deploy_example}: Example of directives under the {\deploy} category}
\vspace{-6pt}
\begin{tabular}{C{7.2cm} | C{7.2cm}}
\multicolumn{1}{c}{\textsf{GitHub~Actions}} & \multicolumn{1}{c}{\textsf{Travis~CI}} \\
\hline
\captionsetup{labelformat=empty}
\lstset{style=yml}
\begin{lstlisting}[label={lst:deploy_example},caption={~},aboveskip=-18pt,belowskip=-8pt]
|<steps>|:
  |<- run>|: |
      |>base64<| --decode > keystore.jks
      |>./gradlew<| publish --no-daemon
    |<env>|:
      |<KEYSTORE_PASSWORD>|: ${{ secrets.PASS }}
\end{lstlisting}
&
\lstset{style=yml}
\begin{lstlisting}[aboveskip=-8pt,belowskip=-8pt]
|<deploy>|:
  |<provider>|: script
  |<script>|: |>./gradlew<| publish
  |<on>|:
    |<branch>|: master
\end{lstlisting}
\\\hline
\end{tabular}
\end{table}

\item \noindent\textbf{\report.} Configuration directives related to sending notifications and generating reports about the CI/CD process. This includes steps to notify team members of build status, test results, and deployment outcomes. Listing~\ref{lst:report_example} gives an example of directives under this category in \textsf{GitHub~Actions} and \textsf{Travis~CI}.

\begin{table}[ht]
\centering
\caption*{Listing~\ref{lst:report_example}: Example of directives under the {\report} category}
\vspace{-6pt}
\begin{tabular}{C{7.2cm} | C{7.2cm}}
\multicolumn{1}{c}{\textsf{GitHub~Actions}} & \multicolumn{1}{c}{\textsf{Travis~CI}} \\
\hline
\captionsetup{labelformat=empty}
\lstset{style=yml}
\begin{lstlisting}[label={lst:report_example},caption={~},aboveskip=-18pt,belowskip=-8pt]
|<steps>|:
  |<- run>|: |>curl<| -X POST 
            -H 'Content-type: application/json' 
            --data '{"text":"Success."}'
            $SLACK_WEBHOOK_URL

\end{lstlisting}
&
\lstset{style=yml}
\begin{lstlisting}[aboveskip=-8pt,belowskip=-8pt]
|<after_success>|:
  |>- curl<| -X POST
         -H 'Content-type: application/json' 
          --data '{"text":"Success."}'
          $SLACK_WEBHOOK_URL
\end{lstlisting}
\\\hline
\end{tabular}
\end{table}

\end{enumerate}

\subsection{Comparisons with CI/CD Configuration Practices in General}
We conduct a thorough review of findings reported in prior studies on CI/CD configuration practices across both open-source and industrial projects, regardless of application domain. To do this, we begin with a recent systematic literature review (SLR) on build optimization~\cite{aidasso2025build} as a base reference, and then perform further cross-referencing to identify studies that specifically address CI/CD configuration. This approach enables us to establish meaningful comparisons with our findings on Android apps and to contextualize the patterns observed in Android-specific pipelines against broader trends and challenges. By examining prior research on configuration smells, anti-patterns, and service-specific practices, we aim for our analysis to capture both domain-specific characteristics and generalizable insights into CI/CD configuration.

\vspace{5pt}
\section{Results} \label{sec:Results}
In this section, we present the results of our research questions (RQs). We discuss the motivation, approach, and findings for each research question.

\subsection{RQ1: How do CI/CD configurations vary in complexity and tool usage for Android apps?}

\subsubsection{\textbf{Motivation.}}
CI/CD practices have become integral to modern software development, facilitating rapid and reliable delivery of software products. However, as CI/CD pipelines grow in complexity and diversity, understanding the complexities of configuration practices and tool usage across different phases of these pipelines becomes critical for several reasons. Firstly, it allows organizations to optimize their CI/CD workflows by identifying potential challenges or inefficiencies in configuration practices. Secondly, it provides insights into the prevalent tools and technologies used by practitioners, enabling informed decision-making regarding tool adoption and integration. Moreover, by shedding light on the disparities in configuration complexity and tool usage, this RQ contributes to the broader discourse on CI/CD best practices and service comparisons. It provides stakeholders, including software developers, CI/CD engineers, and project managers, with actionable insights to enhance the efficiency and effectiveness of their CI/CD processes.

\subsubsection{\textbf{Approach.}}
To address this RQ, we conduct a thorough analysis of CI/CD configurations, focusing on the complexity across different services and categories. We also identify the common tools developers use to test and deploy their mobile apps.

\vspace{3pt}
\noindent\textbf{Analyzing CI/CD configuration complexity.}
We investigate the complexity of configuration files among various CI/CD services to recognize differences. Our analysis primarily focuses on providing descriptive statistics, such as median, maximum value, minimum value, and other relevant measures.
Moreover, to identify significant differences between the complexities of CI/CD configurations of different CI/CD services, we utilize the Wilcoxon test~\cite{wilks2011statistical}. This non-parametric test is particularly suited for comparing two paired groups, which aligns well with our objective of evaluating differences among the services.
We also use the Kruskal–Wallis test~\cite{tomczak2014need} to compare groups, and estimate the effect size using eta squared ($\eta^2$) and report the Kruskal–Wallis statistic ($H$), following recommendations for non-parametric data. This provides an interpretable measure of the proportion of variance explained by the grouping variable.
Finally, we analyze the complexity of directives under different categories to understand which categories require more configuration and attention across various services. To do this, we calculate the number of directive/value pairs under each category for each project independently. Given that configuration size varies across projects, we calculate the percentage of each category for each project. We use boxplots to visualize these percentages, highlighting the descriptive statistics overall and for each CI/CD service.

\vspace{3pt}
\noindent\textbf{Identifying testing and deployment tools.}
We perform a semi-automated approach combining manual and automated analysis in the same processes as in the ``Remove toy projects'' step in Figure~\ref{fig:data_collection} to investigate and analyze the tools used during the deployment and testing phases. This involves examining the usage patterns of various commonly used testing and deployment tools across different CI/CD services, allowing us to understand the prevalence and effectiveness of these tools within the context of CI/CD practices.

\subsubsection{\textbf{Findings.}} We present our findings on CI/CD configuration complexity at different levels: overall, as well as across various CI/CD services, configuration categories, and programming languages.

\vspace{5pt}
\noindent\textbf{Observation 1.1: }\textbf{CI/CD configuration is not standardized, and its complexity varies across CI/CD services, with a median of 19 configuration directives per file, as we observe that \textsf{GitLab~CI/CD} has the highest value while \textsf{Travis~CI} service has the lowest complexity level.}
Our analysis of configuration directives uncovers significant differences in file complexities among CI/CD services, indicating variations in the utilization of configuration directives across these services. The median number of directives per file is $19$, with distinctive differences observed across CI/CD services. Specifically, \textsf{GitLab~CI/CD} stands out with a higher directive count of $25.5$, whereas \textsf{Travis~CI} shows a lower count of $16$. Other services fall in between, with \textsf{GitHub~Actions} at $20$, and \textsf{CircleCI} at $18$.

Our results reveal that, overall, all CI/CD services have high variability in complexity, with many outliers indicating some configuration files are significantly more complex. This implies that those services offer a lot of flexibility and functionality for configuring CI/CD pipelines, but they also require more attention and expertise to manage the complexity. Some configuration files may be unnecessarily complex or inefficient, and could benefit from simplification or optimization.

Our results reveal a significant difference (\textit{p-value} < 0.0001) in the complexities of CI/CD configuration files across various programming languages ($H$ = 429.06, $\eta^2$ = 0.083), indicating a \textbf{\textit{moderate effect size}} and suggesting that at least one programming language exhibits a substantial difference in complexity compared to others.  Furthermore, we observe a significant difference (\textit{p-value} < 0.0001) in the complexity of files across CI/CD services ($H$ = 160.87, $\eta^2$ = 0.032), with a \textbf{\textit{small to moderate effect size}}, highlighting meaningful variations in complexity between different CI/CD services. Therefore, developers should make informed decisions about which CI/CD service to adopt, carefully considering the balance between the required features and the potential complexity of initializing and maintaining the configuration files.

\vspace{4pt}
\noindent
    \textbf{\textit{Compare \& Contrast with Prior Research Findings (1.1):}}
    Overall, our results align with prior research indicating that CI configurations were deemed complex and generally challenging for developers, not only in Android apps~\cite{rostami2023usage,widder2019conceptual,hilton2017trade,kavaler2019tool,zhang2018one,saroar2023developers}. For example, \citet{zhang2018one} indicated that developers can become frustrated with specific CI/CD services and discouraged from adopting CI practices, even when more customized options are available. Some studies, based on developer feedback, linked such complexity to \yml's syntax for being confusing and not easy to write configurations with it~\cite{saroar2023developers}. The lack of standardization across different CI/CD services was also reported~\cite{rostami2023usage}, which made migration between services difficult and error-prone, as developers must navigate differing semantics and feature implementations. There is a learning curve in understanding all the configuration options, and it can be complex to configure even a simple CI pipeline~\cite{hilton2017trade}.
    In the context of our study, Android apps can often require more complex build setups, platform-specific dependencies, and integration with mobile-specific tools (e.g., emulators), making their CI/CD configuration worth investigating across various services.

\begin{figure}
    \centering
        \begin{minipage}{1\textwidth}
            \includegraphics[width=1\linewidth]{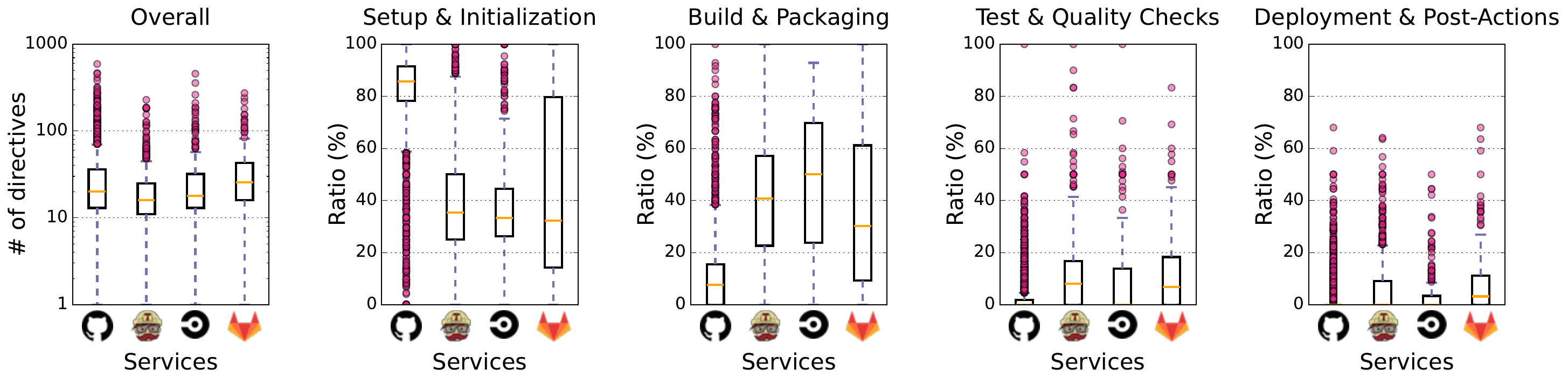}
            \vspace{-15pt}
            \caption{Boxplots showing the total number of directives across CI/CD services, alongside the respective ratios of directives for each category.
            \textbf{Icons:}\hspace{3pt}\raisebox{-0.35ex}{\includegraphics[height=8pt]{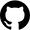}} \textsf{GitHub~Actions} \hspace{3pt}\textbullet\hspace{3pt}
            \raisebox{-0.35ex}{\includegraphics[height=8pt]{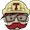}} \textsf{Travis~CI} \hspace{3pt}\textbullet\hspace{3pt}
            \raisebox{-0.35ex}{\includegraphics[height=8pt]{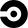}} \textsf{CircleCI} \hspace{3pt}\textbullet\hspace{3pt}
            \raisebox{-0.35ex}{\includegraphics[height=8pt]{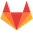}} \textsf{GitLab~CI/CD}}
            \label{fig:rq1-bloxplots}
        \end{minipage}
    \vspace{-13pt}
\end{figure}

\vspace{3pt}
\noindent\textbf{Observation 1.2:}
\textbf{The majority of CI/CD configuration efforts are dedicated to {\setup}, indicating the significance of preparing the environment for successful build and testing processes, with an overall median of $81.3\%$ of the directives in \yml files.}
Figure~\ref{fig:rq1-bloxplots} shows boxplots depicting the total number of directives in \yml files for each CI/CD service, alongside the respective ratios of directives for each category.
Overall, we observe a prevalence of {\setup}, which indicates the critical effort developers dedicate to this phase, emphasizing the need for a robust foundation before proceeding with subsequent CI/CD phases. This could be attributed to the diverse development environments and dependencies associated with Android apps, requiring detailed configuration in this initial phase for smooth pipeline execution. For CI/CD tool builders and Android app maintainers, understanding the significance of {\setup} is crucial. This implies that providing robust support and features for configuring environments and initializations should be a priority. In addition, documentation and best practices for this phase may be beneficial in reducing the onboarding process for developers working with Android projects.

\textsf{GitHub~Actions} configuration files have the highest complexity in {\setup} with a median of $85.71\%$ of the directives. While this indicates how important setting up CI/CD is, it also indicates a barrier for software developers and maintainers. Given that \textsf{GitHub~Actions} is currently the most commonly used CI/CD service on \textsf{GitHub} and most projects suffer from complex build setups, tool developers should invest in exploring new techniques to simplify this practice and make it consistent across \textsf{GitHub} projects.

Moreover, though the {\build} phase is important for creating deployable artifacts in Android apps, we observe varying degrees of focus on this phase among CI/CD services. The median percentage of directives under this category is $11.8\%$, largely influenced by \textsf{GitHub~Actions}, which puts minimal emphasis with a median of $7.5\%$. In contrast, \textsf{CircleCI} shows substantial complexity in this category with a median of $50.0\%$, followed by \textsf{Travis~CI} and \textsf{GitLab~CI/CD} with $40.8\%$ and $30.1\%$, respectively. These figures illustrate how each service prioritizes enhancing the building process for greater efficiency and reliability, reflecting varying focuses on creating deployable artifacts.
Listing~\ref{lst:actions_vs_travis_example} gives an example of how CI/CD configuration in \textsf{GitHub~Actions} (the \textit{AndroidKeyLogger}\footnote{\url{https://github.com/gokulrajanpillai/AndroidKeyLogger}}project) can be more complex than that of \textsf{Travis~CI} (the \textit{BOI-Balance-Checker}\footnote{\url{https://github.com/LukeHackett12/BOI-Balance-Checker}} project) for running an almost similar workflow (i.e., running a \textit{Gradle} build).
Furthermore, we observe that directives from other categories are used with much lower percentages. For example, despite their critical roles in CI/CD pipelines, the {\test} and {\deploy} categories have less complexity compared to the other categories. While this may suggest simplicity among these processes, it may also suggest potential areas of underemphasis by developers. Researchers are encouraged to assess how these phases are managed and whether their simplicity or complexity is associated with historical build success or failure.

\vspace{4pt}
\noindent
    \textbf{\textit{Compare \& Contrast with Prior Research Findings (1.2):}}
    Prior research indicated that the primary focus of CI configuration in software projects is on environments and containers setup~\cite{zhang2022buildsonic,gallaba2018use}. Testing configuration, however, was observed by \citet{rzig2022characterizing} to be prevalent across a large number of projects. Conversely, \citet{gallaba2018use} highlighted that, while deployment configuration is less common, it can demand considerable configuration code when implemented. Though we expect testing and deployment to have more adoption and complexity in Android apps, due to the variety of platforms requiring testing and the complexities of app store deployments, our findings suggest otherwise.

\begin{table}[ht]
\centering
\vspace{-4pt}
\caption*{Listing~\ref{lst:actions_vs_travis_example}: Example of the configuration complexity of \textsf{GitHub~Actions} compared to \textsf{Travis~CI}}
\vspace{-4pt}
\begin{tabular}{C{7.2cm} | C{7.2cm}}
\multicolumn{1}{c}{\textsf{GitHub~Actions}} & \multicolumn{1}{c}{\textsf{Travis~CI}} \\
\hline
\captionsetup{labelformat=empty}
\lstset{style=yml}
\begin{lstlisting}[label={lst:actions_vs_travis_example},caption={~},aboveskip=-18pt,belowskip=-8pt]
|<name>|: Android CI
|<on>|:
  |<push>|:
    |<branches>|: [ master ]
  |<pull_request>|:
    |<branches>|: [ master ]
|<jobs>|:
  |<build>|:
    |<runs-on>|: ubuntu-latest
    |<steps>|:
    |<- uses>|: actions/checkout@v2
    |<- name>|: set up JDK 11
      |<uses>|: actions/setup-java@v2
      |<with>|:
        |<java-version>|: '11'
        |<distribution>|: 'temurin'
        |<cache>|: gradle
    |<- name>|: Grant execute permission for gradlew
      |<run:>| |>chmod<| +x gradlew
    |<- name>|: Build with Gradle
      |<run>|: |>./gradlew<| build
\end{lstlisting}
&
\lstset{style=yml}
\begin{lstlisting}[aboveskip=-8pt,belowskip=-8pt]
|<language>|: android
|<before_install>|:
  - yes | sdkmanager "platforms;android-28"
|<script>|:
  |>- ./gradlew<| build
\end{lstlisting}
\\\hline
\end{tabular}
\vspace{-10pt}
\end{table}

\vspace{3pt}
\noindent\textbf{Observation 1.3:}
\textbf{Nearly half of the projects implement automated testing processes, with the predominant standard testing tool, through \textit{Gradle}, accounting for about 78\% of the total.}
Among the analyzed projects, 1,223 (47.7\%) use automated testing processes as part of the CI/CD pipeline.
Table~\ref{testing_tools} shows the top 10 testing tools used by the projects in our dataset. We should note that multiple testing tools can be used in a single project, resulting in cumulative percentages that exceed 100\%.
The results show a preference for adopting testing practices through standard tools like \textit{Gradle} in Android projects. The results are consistent across CI/CD services. This trend likely arises from the ease of integration and familiarity, leading to more efficient testing workflows.
We observe that unit testing is by far the most commonly used practice, with 78.3\% of projects incorporating it. This highlights the importance of unit testing in software development, as it enables developers to verify code correctness at a granular level. This suggests that unit testing is deemed essential for maintaining code reliability and is likely prioritized due to its effectiveness in identifying defects early and its seamless integration with continuous integration CI/CD pipelines.

Android-specific tools, such as \textit{Android Lint} (28.7\%), testing on connected devices (14.5\%), and the \textit{Android Emulator} (9.6\%), show a relatively lower adoption compared to unit testing, despite their role in ensuring platform compliance and optimizing performance. The greater reliance on \textit{Android Lint} suggests an interest of Android developers in performing static code analysis to detect issues before runtime. However, the low adoption rates of connected devices and emulators might indicate that many teams prefer more lightweight and automated approaches, possibly relying on simulated environments or static checks before moving to the next steps.

The adoption of tools focused on code quality, code coverage, and static analysis, such as Jacoco (5.7\%), Detekt (5.2\%), and CheckStyle (1.1\%), is limited. These tools play a critical role in enforcing code quality and monitoring coverage, which are essential for long-term maintainability. However, their lower usage might indicate that these practices are either under-prioritized or selectively applied, perhaps depending on the project's complexity or the organization's quality assurance strategy. This trend suggests that, while these tools are valued, they are often deemed secondary and may be adopted only in more mature environments.

Finally, tools like \textit{Firebase}\footnote{\url{https://firebase.google.com}} (5.1\%), \textit{Fastlane}\footnote{\url{https://fastlane.tools}} (2.2\%), and \textit{SonarQube}\footnote{\url{https://www.sonarsource.com}} (2.1\%) show even lower adoption rates, despite their popularity as third-party testing platforms. Firebase's usage likely reflects a focus on cloud-based testing and monitoring, particularly for mobile apps, but its usage remains relatively limited in our datasets. \textit{Fastlane} and \textit{SonarQube}, which facilitate CI/CD automation and comprehensive code quality analysis, respectively, appear to be more specialized tools that are leveraged in specific cases, possibly due to their integration complexity.

These findings indicate that, while unit testing remains a staple across Android projects, there is a gap in the widespread adoption of advanced code quality and static analysis tools. Hence, software developers are encouraged to enhance code reliability by integrating more comprehensive quality checks. Researchers should explore the factors limiting the broader adoption of these tools among Android projects and the impact on overall software quality. CI/CD services should also consider offering more user-friendly integrations for less-used tools, such as Jacoco and SonarQube, promoting their broader adoption and optimizing automated testing workflows.

\begin{table}[ht]
\centering
\caption{Top 10 testing tools adopted in our dataset}
\label{testing_tools}
\begin{tabular}{ l l L{9cm} l r }
\toprule
  & \textbf{Testing Tool} & \textbf{Configuration Examples} & \# projects (\%) \\
\midrule
1  & Unit testing            & \texttt{./gradlew test | check | testDebug | testDebugUnitTest |}  & 958 (78.3\%) \\
   &                         & \texttt{testReleaseUnitTest}                                       &              \vspace{3pt}\\
2  & Android Lint            & \texttt{./gradlew lint | klint | ktlintcheck | ktlintformat}       & 351 (28.7\%) \vspace{3pt}\\
3  & Testing on connected    & \texttt{./gradlew connectedcheck | connectedAndroidTest |}         & 177 (14.5\%) \\
   & devices                 & \texttt{connectedDebugAndroidTest | connectedFullDebugAndroidTest} &              \vspace{3pt}\\
4  & Android Emulator        & \texttt{uses: reactivecircus/android-emulator-runner@v2}           & 118 (9.6\%) \\              
5  & Jacoco                  & \texttt{./gradlew jacocotestreport}                                & 70  (5.7\%) \\
6  & Detekt                  & \texttt{./gradlew detekt}                                          & 64 (5.2\%) \\
7  & Firebase                & \texttt{gcloud firebase test android run}                          & 62 (5.1\%) \\        
8  & Fastlane                & \texttt{fastlane android test}                                     & 27 (2.2\%) \\        
9  & SonarQube               & \texttt{./gradlew sonarqube}                                       & 26 (2.1\%) \\        
10 & CheckStyle              & \texttt{./gradlew checkstyle}                                      & 14 (1.1\%) \\        
\bottomrule
\end{tabular}
\end{table}

Listing~\ref{lst:testing_tools_examples} gives three examples of how different testing and quality assurance tools are used in CI/CD configurations in \textsf{GitHub~Actions}, namely \textit{Gradle Unit Testing}, \textit{Android Lint}, and \textit{Android Emulator}, demonstrating how complex it is to test apps on an emulator compared to standard \textit{Gradle} testing.

\vspace{4pt}
\noindent
    \textbf{\textit{Compare \& Contrast with Prior Research Findings (1.3):}}
    Similar to our findings, \citet{zhang2022buildsonic} indicated that \textit{Gradle} is a prevalent testing tool. However, \textit{Maven}, which was observed to be dominant in Java projects~\cite{zhang2022buildsonic}, appears to be less common for Android apps. In addition, \citet{rzig2022characterizing} reported various tools for testing machine learning projects developed with Python, underscoring the need to investigate Android-specific testing tools. We discover several tools unique to Android testing (e.g., \textit{Android Lint} and \textit{Android Emulator}) that are not commonly used in regular software testing.

\begin{table}[ht]
\centering
\caption*{Listing~\ref{lst:testing_tools_examples}: Examples of how different testing tools are used in CI/CD configurations (\textsf{GitHub~Actions})}
\vspace{-6pt}
\begin{tabular}{C{3.4cm} | C{3.55cm} | C{7.06cm}}
\multicolumn{1}{c}{Gradle Unit Testing} & \multicolumn{1}{c}{Android Lint} & \multicolumn{1}{c}{Android Emulator} \\
\hline
\captionsetup{labelformat=empty}
\lstset{style=yml}
\begin{lstlisting}[label={lst:testing_tools_examples},caption={~},aboveskip=-18pt,belowskip=-8pt]
|<test>|:
 |<- name>|: Run Unit Tests
 |<- run>|: |>./gradlew<| test
\end{lstlisting}
&
\captionsetup{labelformat=empty}
\lstset{style=yml}
\begin{lstlisting}[caption={~},aboveskip=-18pt,belowskip=-8pt]
|<test>|:
 |<- name>|: Run Android Lint
 |<- run>|: |>./gradlew<| lint
\end{lstlisting}
&
\multirow{2}{*}{}
\captionsetup{labelformat=empty}
\lstset{style=yml}
\begin{lstlisting}[aboveskip=-8pt,belowskip=-8pt]
|<test>|:
 |<- name>|: Run Android instrumentation tests
  |<uses>|: reactivecircus/android-emulator-runner@v2
\end{lstlisting}
\\\cline{1-2}
\multicolumn{2}{c}{}
\captionsetup{labelformat=empty}
\lstset{style=yml}
\begin{lstlisting}[caption={~},aboveskip=-18pt,belowskip=-8pt]
|<test>|:
 |<- name>|: Run Unit Tests and Lint
 |<- run>|: |>./gradlew<| test lint
\end{lstlisting}
\hfill\vline
&
\captionsetup{labelformat=empty}
\lstset{style=yml}
\begin{lstlisting}[aboveskip=-22pt,belowskip=-8pt]
  |<with>|:
    |<api-level>|: 29
    |<script>|: |>./gradlew<| connectedCheck
\end{lstlisting}
\\\hline
\end{tabular}
\end{table}

\vspace{3pt}
\noindent\textbf{Observation 1.4:}
\textbf{Among the studied projects, only 9\% manage deployments, using diverse methods like \textsf{GitHub} services, external scripts, and third-party services.}
Our results reveal 238 projects managing deployments as part of their CI/CD pipelines.
Table~\ref{deployment_tools} shows the top 10 deployment tools used by the projects in our dataset.
We observe that only 27 (11\%) deploy directly to the \textsf{Google Play Store}, whereas the majority (89\%) employ third-party tools or platforms for deployment, suggesting a preference for external tools over direct \textsf{Google Play Store} integration, likely due to potential barriers (e.g., complexity or limitation) in direct deployment practices.

\begin{table}
\centering
\caption{Top 10 deployment tools adopted in our dataset}
\label{deployment_tools}
\begin{tabular}{ c l r r r r r }
\toprule
   & \textbf{Deployment Tool} & \textsf{\textbf{GitHub~Actions}} & \textsf{\textbf{Travis~CI}} & \textsf{\textbf{CircleCI}} & \textsf{\textbf{GitLab~CI/CD}} & \textbf{Total} \\
\midrule
    1 & GitHub & 13 & 91 & 4 & 0 & 118 \\
    2 & Script & 33 & 12 & 4 & 3 & 52 \\
    3 & Fastlane & 32 & 1 & 6 & 5 & 44 \\
    4 & Google Play & 21 & 1 & 3 & 2 & 27 \\
    5 & SSH & 8 & 1 & 1 & 3 & 13 \\
    6 & SonaType & 14 & 0 & 0 & 0 & 14 \\
    7 & S3 & 3 & 2 & 0 & 4 & 9 \\
    8 & Google Cloud & 2 & 1 & 1 & 0 & 4 \\
    9 & FDroid & 4 & 0 & 0 & 1 & 5 \\
    10 & Semantic-release & 2 & 0 & 0 & 2 & 4 \\
\bottomrule
\end{tabular}
\end{table}

Deployment practices vary among developers. Notably, 118 (50\%) projects use \textsf{GitHub} tools, such as \textit{Releases} and \textit{Pages}, while 52 (22\%) projects use external scripts (e.g., \texttt{shell} scripts). We also observe that \textit{Fastlane} is used by 44 (18\%) projects, whereas only a few projects are found to be deployed on unofficial platforms, such as \textsf{F-Droid} (5 projects).
These results suggest that developers may face challenges in automating app deployments to the \textsf{Google Play Store}, as the most popular \textsf{GitHub~Actions} template recommends manually uploading an \textit{APK} or \textit{AAB} file via the Play Console\footnote{\url{https://github.com/marketplace/actions/upload-android-release-to-play-store}}. However, deployment through the console requires APKs to be signed by the developer's private key, which ensures developer's authenticity, but still adds an extra layer of complexity to the process. This highlights the need for better automation solutions to deploy to the \textsf{Google Play Store} directly through \textsf{GitHub~Actions}.
\textsf{Travis~CI} projects rarely automate direct deployments to the \textsf{Google Play Store}---only one project does. Instead, 91 out of 105 projects use other services, such as \textsf{GitHub Pages}, likely to host documentation, release notes, or web-based demos of their apps, revealing distinct deployment habits compared to projects that adopt \textsf{GitHub~Actions}.
For \textsf{CircleCI}, most users face hurdles and resort to third-party solutions like \textit{Fastlane}, with only 18 projects including deployment automation and three deploying directly to the \textsf{Google Play Store}.
\textsf{GitLab~CI/CD} users automate deployments in 13 projects, mainly using \textit{Fastlane} (five projects), as well as AWS S3, external scripts, and SSH for web hosting. Only two projects deploy directly to the \textsf{Google Play Store}.

Listing~\ref{lst:deploy_tools_examples} provides two examples of how deployment tools are adopted in CI/CD. One example uses \textit{Fastlane} with \textsf{Travis~CI}, while the other showcases direct deployment to the \textsf{Google Play Store} with \textsf{GitHub~Actions}. These examples illustrate the complexity of configuring deployments in CI/CD environments, given the large number of parameters involved. Moreover, some of these tools are proprietary and may require payment. This indicates diverse deployment preferences and challenges across CI/CD services, suggesting the need for more integrated and automated solutions to streamline mobile app deployments to popular platforms like the \textsf{Google Play Store}.
These inconsistencies in deployment practices can affect reproducibility, maintainability, and contributor onboarding. For example, tool-specific configurations may introduce integration overheads or failures that are hard to debug and replicate. Inconsistent automation strategies can also complicate release planning under the typical rapid development cycles of Android apps. Ultimately, the lack of standardized deployment configurations hinders the uniform adoption of quality assurance practices across projects.

\vspace{4pt}
\noindent
    \textbf{\textit{Compare \& Contrast with Prior Research Findings (1.4):}}
    \citet{gallaba2018use} and \citet{rzig2022characterizing} reported low adoption rates for deployment in Java and Python projects, respectively, with relatively higher rates in Python. Though some CI/CD services like \textsf{Travis~CI} offer built-in deployment support \cite{rzig2022characterizing}, the authors found many third-party tools being used for deployment. These tools tend to deploy to cloud services, like \textsf{AWS} or \textsf{Azure}, rather than specific app stores. Moreover, \citet{gallaba2018use} noted that organizations might avoid using the deployment phase in CI due to the potential complexity of its configuration. In the context of Android apps, we also identify several third-party tools used for deployment, but with a noticeable preference for deploying to GitHub, as uploading to app stores is often performed manually. This observation contradicts a prior finding on industrial projects, which advocated for automated deployment over manual methods to accelerate release cycles and minimize errors~\cite{DBLP:journals/corr/abs-2103-04251}.

\begin{table}[ht]
\centering
\caption*{Listing~\ref{lst:deploy_tools_examples}: Examples of how different deployment tools are used in CI/CD configurations}
\vspace{-6pt}
\begin{tabular}{C{7.2cm} | C{7.2cm}}
\multicolumn{1}{c}{Fastlane (\textsf{Travis~CI})} & \multicolumn{1}{c}{Directly to Google Play (\textsf{GitHub~Actions})}\\
\hline
\captionsetup{labelformat=empty}
\lstset{style=yml}
\begin{lstlisting}[label={lst:deploy_tools_examples},caption={~},aboveskip=-18pt,belowskip=-8pt,escapechar=\!]
|<deploy>|:
 |<- provider>|: script
   |<script>|: |>bundle exec fastlane supply<|
           |>run<| --apk aw-android*.apk
   |<skip_cleanup>|: true
   |<on>|:
     |<tags>|: true
     |<repo>|: ActivityWatch/aw-android
 |<- provider>|: releases
   |<file>|: aw-android*.apk
   |<file_glob>|: true
   |<skip_cleanup>|: true
   |<on>|:
     |<tags>|: true
     |<repo>|: ActivityWatch/aw-android
   |<api_key>|:
     |<secure>|: <key> ! {\color{codegreen}\# key is hidden}!

\end{lstlisting}
&
\lstset{style=yml}
\begin{lstlisting}[aboveskip=-8pt,belowskip=-8pt]
|<jobs>|:
 |<UploadAABFileToPlayStore>|:
  |<runs-on>|: ubuntu-latest
  |<steps>|:
   ...
   |<- name>|: Upload AAB to Google Play Store
     |<uses>|: r0adkll/upload-google-play@v1
     |<with>|:
       |<serviceAccountJson>|: service_account.json
       |<packageName>|: com.beomjo.whitenoise
       |<releaseFiles>|: .../release/*.aab
       |<track>|: internal
       |<inAppUpdatePriority>|: 5
\end{lstlisting}
\\\hline
\end{tabular}
\end{table}

\begin{Summary}{}{}
CI/CD configurations are not standardized and vary in complexity across categories and CI/CD services. Most efforts focus on pipeline setup, comprising over 81\% of directives. Testing is performed in nearly half of the projects, with a predominant focus on standard unit testing, whereas deployment is infrequent, occurring in only 9\% of the projects.
These findings align to some extent with prior research findings on general-purpose projects, though our results contrast in showing lower-than-expected adoption of testing and deployment in Android apps despite their inherently demanding release requirements.
\end{Summary}

\vspace{4pt}
\textbf{\subsection{RQ2: How do CI/CD configurations evolve in Android apps?}}

\subsubsection{\textbf{Motivation.}}
Understanding the evolution of CI/CD configuration is important as it reveals how CI/CD pipelines mature and adapt over time. By examining trends in configuration changes, this RQ aims to identify how CI/CD configurations evolve to either integrate new features or resolve existing issues. Analyzing this evolution can help software developers understand the importance of periodically maintaining CI/CD pipelines to enhance their efficiency and effectiveness.

\subsubsection{\textbf{Approach.}}
In this RQ, we collect and process the commits in which developers change CI/CD configuration files to understand the complexity of changes, evolution trends, and factors associated with change frequency.

\vspace{4pt}
\noindent\textbf{Analyzing complexity and evolution of changes.}
To analyze the evolution of CI/CD configuration complexity, we segment the data by project and commit period (monthly).
To visualize the number of changes (commits) per project across CI/CD services, we use boxplots in which we display commit distribution overall and per CI/CD services, showing variations and outliers.
To analyze the evolution of changes across various CI/CD services over the years, we start by identifying the first commit for each project to grasp the initial configuration state. We then calculate the total number of changing commits for each project. These counts are normalized by the total number of active projects adopting CI/CD within a given year, accounting for the newly adopting projects, as follows:
\[
\text{Average \# of commits per project} = \frac{\sum_{i=1}^{n} C_i}{R}
\]
where \( C_i \) is the number of commits for the \( i^{th} \) project, \( n \) is the total number of projects, and \( R \) is the number of active projects in a given year. We use line plots to visualize trends in CI/CD maintenance activity by illustrating the average number of commits per project over the years for each CI/CD service.

\vspace{4pt}
\noindent\textbf{Analyzing CI/CD evolution trends.}
In addition, we perform linear regression to calculate a trend coefficient from (a) the total number of commits, and (b) the number of lines added and removed (i.e., modified) over time for each project. These coefficients are then classified into \textit{positive}, \textit{negative}, or \textit{steady} trends.
We also calculate statistics that capture the overall patterns of CI/CD configuration changes for each category, revealing distinct trends in increasing, decreasing, or stable complexity across projects. This approach offers a clear, systematic view of how CI/CD practices evolve in various software development contexts.
In addition, when we analyze configuration changes across categories, we identified new sets of directives and values that did not appear in our original analysis. Therefore, we categorize the newly identified pairs of directives and values using the heuristics described in Section~\ref{categorizing_directives}. However, we find a small portion of ambiguous pairs (less than $2\%$) in which the directives do not belong to any of the categories, which we discard from our analysis.

\vspace{4pt}
\noindent\textbf{Understanding factors associated with CI/CD change frequency.} To understand the factors associated with CI/CD maintenance activity in software projects, by identifying the significant variables associated with the number of configuration commits in the studied projects using linear regression. To do this, we fit an Ordinary Least Squares (OLS) model, a common method for estimating the parameters in a linear regression model, identifying the relationships between variables. To fit our model, we use common project characteristics provided by \textsf{GitHub} about software repositories, as shown in Table~\ref{tab:model_vars}.
Specifically, we use the \textit{number of configuration commits} as the dependent variable and the other project characteristics as independent variables. We assess the performance of the model using the adjusted $R^2$, which measures the proportion of variance in the dependent variable explained by the independent variables. The adjusted $R^2$ helps to adjust the obtained $R^2$ value for the number of predictors, thus preventing overfitting when using multiple variables. An $R^2$ value closer to $1.00$ indicates a better fit. We also evaluate the significance of each factor based on the p-values produced by the model, and identify the directionality of their relationships through coefficient analysis (i.e., either positively associated (\up) or negatively associated (\dn)).

\begin{table}
\centering
\caption{Variables used to model the number of configuration commits}
\vspace{-7pt}
\label{tab:model_vars}
\begin{tabular}{p{4.4cm} p{10cm}}
\toprule
    \textbf{Variable} & \textbf{Description} \\
\midrule
    Repo Size & The total size of the repository, typically measured in kilobytes \\
    Repo Age & The age of the repository in days, from the dates of the first and last activity \\
    Repo Stars & The number of stars received by the repository, indicating its popularity \\
    Repo Watchers & The number of followers of the repository \\
    Repo Forks & The number of times the repository has been forked \\
    Adopting \textsf{GitHub~Actions} & A binary indicator of whether the repository adopts \textsf{GitHub~Actions} \\
    Adopting \textsf{Travis~CI} & A binary indicator of whether the repository adopts \textsf{Travis~CI} \\
    Adopting \textsf{CircleCI} & A binary indicator of whether the repository adopts \textsf{CircleCI} \\
    Adopting \textsf{GitLab~CI/CD} & A binary indicator of whether the repository adopts \textsf{GitLab~CI/CD} \\
    \# of \textsf{Google Play Store} Links & The number links to the \textsf{Google Play Store} \\
    \# of Packages & The total number of application packages \\
    \# CI/CD Services & The total number of CI/CD adopted by the repository \\
\bottomrule
\vspace{-7pt}
\end{tabular}

\end{table}

\vspace{7pt}
\subsubsection{\textbf{Findings.}}
~

\begin{figure}
    \centering
    \includegraphics[width=1\linewidth]{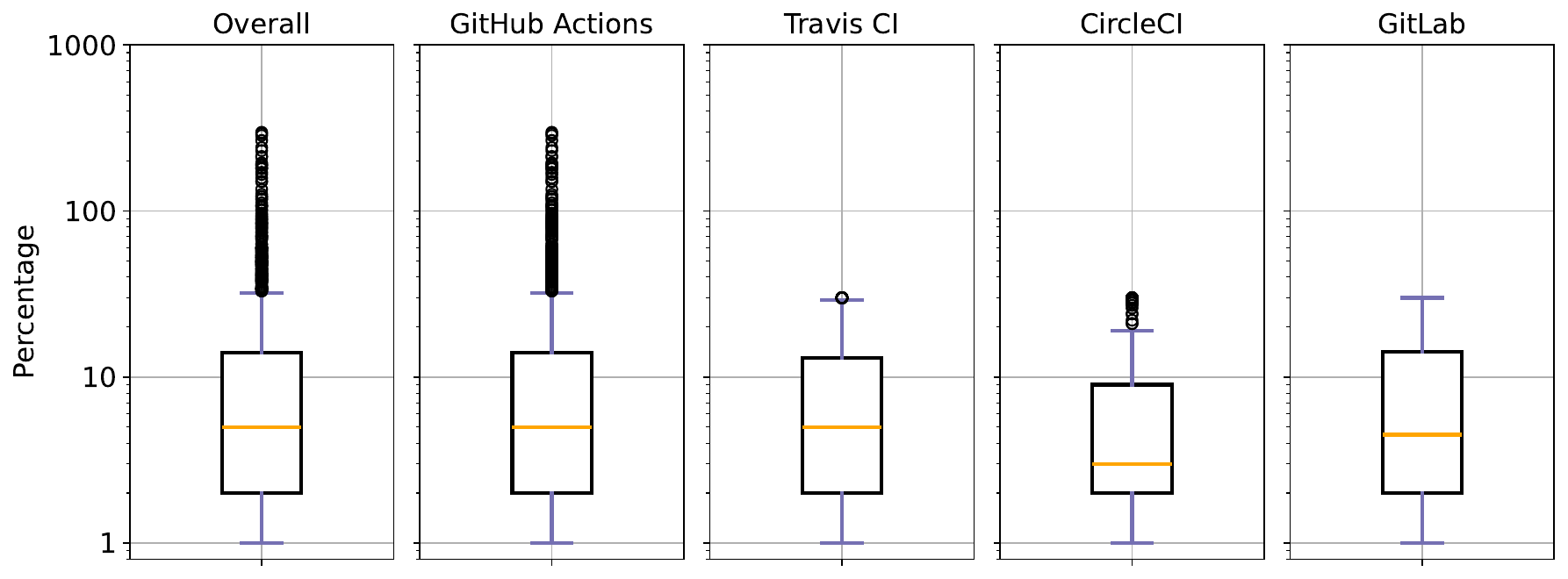}
    \vspace{-15pt}
    \caption{Boxplots showing the number of CI/CD configuration changes (commits), overall and across CI/CD services}
    \vspace{-7pt}
    \label{fig:rq2-bloxplots}
\end{figure}

\noindent\textbf{Observation 2.1:}
\textbf{Developers typically make a median of five commits (about one commit per month) to CI/CD configuration files, with a median of two modifications per commit, indicating regular, incremental updates to CI/CD configurations.}
Figure~\ref{fig:rq2-bloxplots} shows the number of commits changing CI/CD configuration files, overall and across CI/CD services. We observe that projects encounter a median of five commits (mean of 11 commits), indicating a skew towards projects with fewer commits. This variation suggests different CI/CD practices, likely influenced by factors such as project complexity, team size, and development methodologies.
When it comes to commits per project lifetime, we identify 19\% projects with only a single commit per CI/CD service, which indicates that CI/CD was never maintained through the development life cycle. The remaining projects encountered a median of one commit per month, suggesting that a large number of projects have relatively infrequent CI/CD configuration changes, while a few projects with very high commit frequencies.

In terms of actual CI/CD configuration modifications (i.e., adding or removing directives or values), we observe that developers generally make gradual adjustments to CI/CD configurations, with a median of about two modifications per commit.
Nevertheless, we observe other occasional large changes in some projects, likely due to cumulative changes (e.g., merges), indicating some variability in CI/CD practices across projects. 
Moreover, despite a similar average number of changes across various CI/CD services, we observe that \textsf{GitHub~Actions} has significantly more outliers compared to other services, indicating a substantial maintenance overhead associated with this service.
This variability highlights the adaptability of CI/CD practices and emphasizes the importance for teams and tool providers to remain flexible and responsive to changing requirements and opportunities for improvement.

\vspace{4pt}
\noindent
    \textbf{\textit{Compare \& Contrast with Prior Research Findings (2.1):}}
    Prior research on general-purpose projects reported lower rates of change in CI configurations for most projects using \textsf{Travis~CI}~\cite{hilton2016usage,zampetti2021ci}. However, \citet{khatami2024catching} found higher change rates of CI/CD pipelines with \textsf{GitHub~Actions}. Consistent with the former and in contrast to the latter, our study reveals that Android apps exhibit low change rates across all CI/CD services.

\begin{figure}[ht]
    \vspace{-10pt}
    \centering
    \includegraphics[width=1\linewidth]{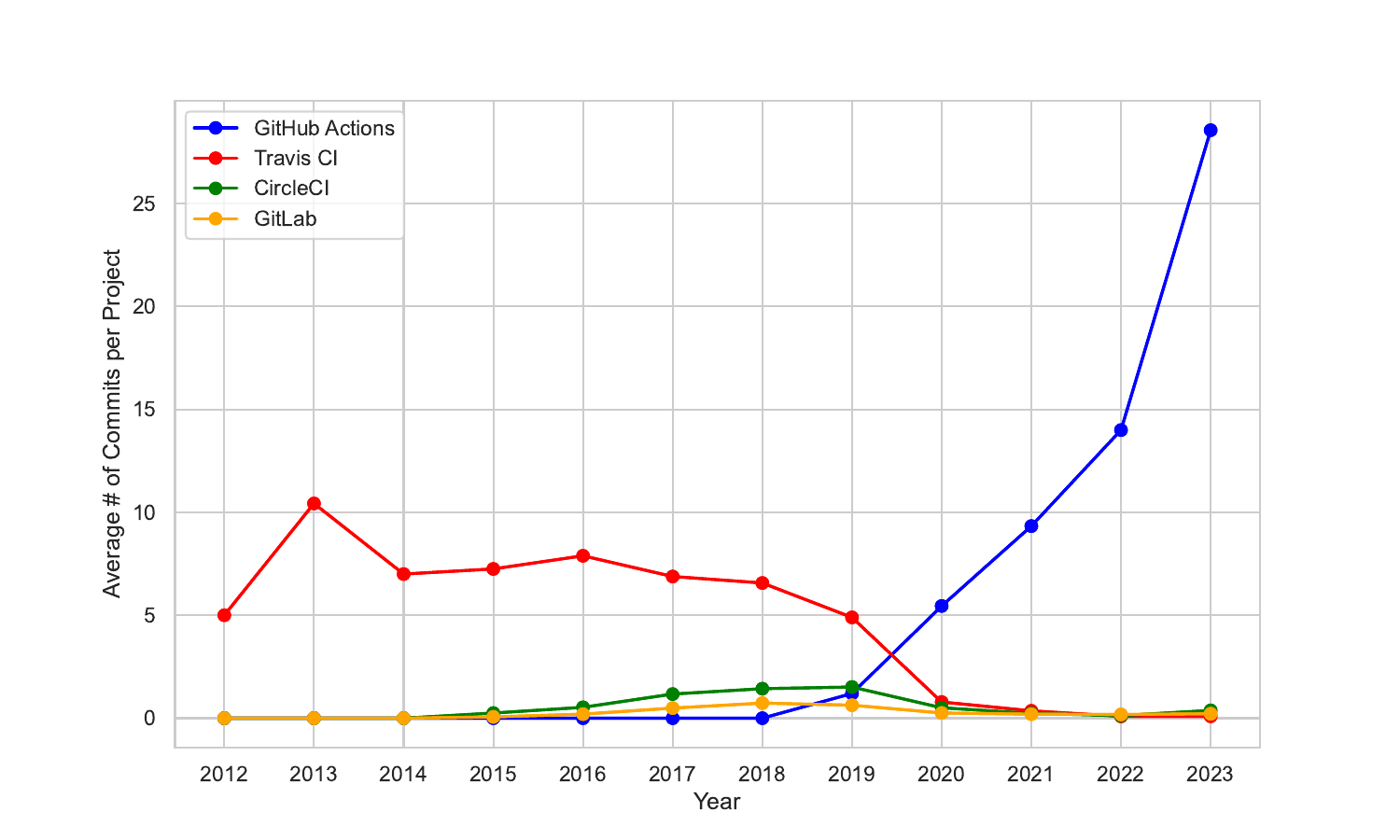}
    \vspace{-20pt}
    \caption{Line plot showing the average number of configuration changes (commits) over time across CI/CD services}
    \label{fig:changes_over_time}
\end{figure}

\vspace{4pt}
\noindent\textbf{Observation 2.2:}
\textbf{The majority of projects (52\%) show a decreasing trend in CI/CD changes (commits) over time, with 28\% and 20\% exhibiting an increasing and steady rate, respectively, of actual configuration lines modified, suggesting simplification or stabilization in CI/CD adoption.}
Figure~\ref{fig:changes_over_time} shows an overall trend of configuration changes (commits) over time across the different CI/CD services. We observe that \textsf{GitHub~Actions} shows a rapid increase in configuration changes, due to shifting towards it after its introduction in 2018. In contrast, \textsf{Travis~CI} exhibits a decline after 2017, nearly reaching zero configuration changes after 2020, likely due to the decreased adoption among \textsf{GitHub} projects due to being less popular for open-source projects. \textsf{CircleCI} and \textsf{GitLab~CI/CD} demonstrate steady growth peaking in 2018-2019, followed by a steady decrease, indicating moderate user retention, likely amidst the introduction of \textsf{GitHub~Actions}.
However, the surge in configuration commits for projects adopting \textsf{GitHub~Actions} may reflect more than mere popularity; it could also suggest potential instability or complexity, prompting frequent adjustments and updates by developers.

Our time series analysis reveals that the majority of projects (52\%) have a negative trend in the number of changes (commits), suggesting a tendency towards simplification or stabilization of CI/CD configurations after initial setup or development phases. Still, a large proportion (28\%) exhibits an increasing trend, which may indicate expanding project requirements or increased adoption of advanced CI/CD features. 20\% of projects maintain a steady level of changes, implying that most projects undergo significant evolution in their CI/CD practices.

With respect to the complexity of configuration changes, we find conflicting trends in the addition and deletion of directives. Specifically, we observe 19\% of projects with an increasing number of changes, 42\% with a decreasing number, and 41\% without significant change. This indicates that many projects are either simplifying their CI/CD setups or maintaining consistent configuration complexity. These trends highlight a dynamic CI/CD landscape, where some projects become more complex or streamlined, while others remain consistent, reflecting differing strategies and stages in project lifecycles.

Moreover, we observe that the frequency of commits has a very weak positive correlation with the number of lines added and deleted, with \textit{Pearson} coefficients of 0.035 and 0.075, respectively. This indicates that there is no strong link between commit frequency and commit complexity. However, there is a moderate correlation between the trends of committed changes and lines added/deleted over time (\textit{Pearson} coefficients of 0.27 and 0.28, respectively). This suggests a general increase in development activity and complexity, with commit frequency moderately linked to code changes.

\vspace{4pt}
\noindent
    \textbf{\textit{Compare \& Contrast with Prior Research Findings (2.2):}}
    \citet{gallaba2018use} observed that most \textsf{Travis~CI} configuration files rarely change once committed. Though \citet{khatami2024catching} observed high volumes of commits modifying \textsf{GitHub~Actions} configurations, there was no indication whether such changes increased or decreased over time.
    \citet{zampetti2021ci} support the fact that developers refactor complex CI/CD configurations or use external scripts to simplify their pipelines.

\vspace{4pt}
\noindent\textbf{Observation 2.3:}
\textbf{CI/CD configurations undergo bi-monthly maintenance (two months, on average), yet many projects have more frequent configuration changes (a median of 2.8 hours), revealing diverse CI/CD maintenance practices.}
Our analysis of intervals between CI/CD configuration changes across various projects reveals a notable disparity: the average time between changes is approximately 61 days, suggesting a bi-monthly update frequency, while the median of about 2.8 hours indicates that many projects have much more frequent updates. This significant difference, coupled with a high standard deviation of about 84 days, suggests a wide range of maintenance practices in CI/CD configurations. This variation reflects the diverse approaches to CI/CD management among different projects, influenced by factors such as project requirements, team dynamics, and development methodologies.

\vspace{4pt}
\noindent
    \textbf{\textit{Compare \& Contrast with Prior Research Findings (2.3):}}
    While prior studies did not emphasize how frequently developers change CI/CD configurations, they reported that such changes are rare. In addition, prior research observed that some projects experience significantly more modifications to \textsf{Travis~CI} configurations than others, aligning with our observations in the context of Android apps across all CI/CD services. Though \citet{ghaleb2019studying,ghaleb2022studying} reported that projects facing build failures or long durations might attempt more frequent configurations to resolve these issues, findings reported by \citet{zampetti2020empirical} suggest that CI/CD maintenance may be intended to address certain configuration smells or anti-patterns.

\vspace{4pt}
\noindent\textbf{Observation 2.4:}
\textbf{For popular CI/CD services, such as \textsf{GitHub~Actions} and \textsf{Travis~CI}, changes focus primarily on {\setup} and {\build}, while changes are infrequent for {\deploy} across all services, indicating stable deployment processes.}
Table~\ref{tab:changes_across_categories_services} summarizes configuration changes across various categories and CI/CD services.
Overall, {\setup} accounts for 69\% of total changes, highlighting its importance in the CI/CD process. {\build} follows with 18\%, suggesting that it is relatively stable once configured. We also observe that {\test} (7\%) and {\deploy} (4\%) are less frequently modified, indicating greater stability in these areas. {\report} (<1\%) reflects a ``set-and-forget'' approach, thus requiring minimal changes once set up. This analysis reveals a gap in continuous improvement for testing and deployment in CI/CD processes, suggesting these areas could benefit from increased focus and innovation.
Across CI/CD services, our analysis of configuration changes in CI/CD services shows: \textsf{GitHub~Actions} emphasizes {\setup} (80\%) over {\build} (12\%) and {\test} (5\%), suggesting higher setup complexity and variability. In \textsf{Travis~CI}, \textsf{CircleCI}, and \textsf{GitLab~CI/CD}, most changes are in {\build} (44\%, 38\%, and 47\%, respectively) followed by {\setup} (33\%, 33\%, and 18\%, respectively) indicating a focus on pipeline maintenance.

\begin{table}[ht]
\centering
\caption{Configuration changes across categories and CI/CD services}
\label{tab:changes_across_categories_services}
\begin{tabular}{ l  r  r  r  r  r }
\toprule
\textbf{Category} & \textbf{Overall} & \textsf{\textbf{GitHub~Actions}} & \textsf{\textbf{Travis~CI}} & \textsf{\textbf{CircleCI}} & \textsf{\textbf{GitLab~CI/CD}} \\
\midrule
\setup  & \textbf{69\%} & \textbf{80\%}    & 33\%          & 33\%          & 18\%           \\
\build  & 18\%          & 12\%             & \textbf{44\%} & \textbf{38\%} & \textbf{47\%}  \\
\test   & 7\%           & 5\%              & 15\%          & 21\%          & 17\%           \\
\deploy & 4\%           & 3\%              & 7\%           & 7\%           & 17\%           \\
\report & 1\%           & 1\%              & 1\%           & 2\%           & 0\%            \\
\bottomrule 
\end{tabular}
\end{table}

These results suggest that software developers should streamline and stabilize {\setup} processes, especially in services like \textsf{GitHub~Actions}. High change frequency in {\build} across services indicates a need for continuous adaptation to new tools and practices. Researchers should investigate the causes of frequent changes and their impact on development efficiency and quality. CI/CD service providers should enhance user experience and feature sets in high-change areas by offering more intuitive setup processes or flexible build configurations. Improved documentation, templates, and automated tools can help developers manage frequent changes. Integrating advanced analytics to suggest optimal configurations could reduce the need for adjustments, leading to more stable and efficient CI/CD pipelines.

Moreover, the analysis of CI/CD configuration changes reveals that {\deploy} are infrequent (e.g., 4\% overall, 3\% in \textsf{GitHub~Actions}, and 7\% in \textsf{Travis~CI}), suggesting a stable deployment process, particularly in Android development. This stability benefits developers with consistent pipelines and minimal maintenance, but also indicates limited adaptation to evolving Android technologies. Researchers should explore the balance between deployment effectiveness and innovation. CI/CD providers should enhance Android deployment features by adding tools, automated testing, and robust rollback options, thus fostering agile development cycles amid changing user expectations and market dynamics.

\vspace{4pt}
\noindent
    \textbf{\textit{Compare \& Contrast with Prior Research Findings (2.4):}}
    Prior research on \textsf{Travis~CI} found that most configuration changes address build job processing nodes or steps~\cite{gallaba2018use,zampetti2021ci}, likely related to setup and build processes. Research on \textsf{GitHub~Actions}, however, discovered that most configuration changes deal with testing and other performance issues \cite{bouzenia2024resource}. Though \citet{khatami2024catching} observed an increase in the adoption of deployment practices in \textsf{GitHub~Actions}, we do not observe active maintenance or improvements in the context of Android apps.

\vspace{4pt}
\noindent\textbf{Observation 2.5:}
\textbf{Active issue tracking, repository size/age, and community engagement are significantly associated with CI/CD configuration maintenance.}
Table~\ref{tab:model_results} presents the results of the OLS linear regression model, expressed by the modeling coefficients, p-values, significance, and relationship direction. Though our model obtains an adjusted $R^2$ value of $0.139$, meaning it explains about 14\% of variability in the number of configuration commits, it still offers valuable insights into some of the key factors associated with CI/CD pipeline maintenance activity.
We observe that the number of reported issues is significantly associated with CI/CD commit frequency, indicating that developers are likely to update CI/CD pipelines to address issues in their projects. Similarly, repository size and age positively correlate with configuration commits, suggesting larger and older repositories attract more CI/CD activity, possibly due to a larger user base or more bugs/features. Furthermore, we find that repository stars and watchers also positively correlate with commit frequency related to CI/CD configurations, making popular projects more active in addressing issues and deploying app updates, which highlights the community engagement's role in development activity. Unlike stars and watchers, we find that repository forks are inversely associated with commit frequency. Forks often represent independent development paths, leading to fewer contributions to the original CI/CD pipelines, thus reducing centralized commit activity.
Our analysis also explores the relationship between CI/CD services and maintenance activity. We find that integrating \textsf{GitHub} Actions is a significant positive contributor to commit frequency. Other CI/CD services might have positive relationships with commit activity, but lack statistical significance, indicating that evidence is not robust enough to reliably confirm their association.
Overall, this analysis highlights the importance of certain factors, such as issue tracking, repository age, community engagement, and adoption of CI/CD practices, in fostering an active development environment, while also identifying potential deterrents to commit activity.

\vspace{4pt}
\noindent
    \textbf{\textit{Compare \& Contrast with Prior Research Findings (2.5):}}
    In regular projects using \textsf{Travis~CI}, \citet{felidre2019continuous} found no correlation between project size and CI change frequency, which contradicts our findings. In terms of project age, \citet{zampetti2021ci} found a positive correlation with the frequency of CI changes, aligning with our results. Though prior studies did not explicitly mention associations between CI changes and issue tracking, some studies reported difficulties in troubleshooting build failures~\cite{widder2019conceptual} and adjustments in CI configurations to address these failures~\cite{ghaleb2022studying}, suggesting a potential link. This suggests that Android apps are distinguished by their unique patterns of CI changes, influenced by factors such as project issue tracking, size, and age.

\begin{table}
    \centering
    \caption{Summary of the linear regression model results showing the most important variables associated with the configuration commit frequency}
    \vspace{-10pt}
    \begin{tabular}{p{3.7cm} r r r}
    \toprule
        \textbf{Variable} & \textbf{Coefficient} & \textbf{P-Value} & \textbf{Significance} \\
    \midrule
        \# of Issues                  & 0.02052 & <0.001 & *** (\up) \\
        Repo Size                     & 0.00003 & <0.001 & *** (\up) \\
        Repo Age                      & 0.00128 & 0.003 & ** (\up) \\
        Repo Stars                    & 0.00068 & 0.006 & ** (\up) \\
        Repo Watchers                 & 0.00068 & 0.006 & ** (\up) \\
        Repo Forks                    & -0.00449 & 0.022 & * (\dn) \\
        Adopting \textsf{GitHub~Actions}       & 13.20505 & 0.026 & * (\up) \\
        Adopting \textsf{Travis~CI}            & 7.85544 & 0.190 &  \\
        Adopting \textsf{CircleCI}             & 7.65946 & 0.211 &  \\
        Adopting \textsf{GitLab~CI/CD}               & 7.13153 & 0.263 &  \\
        \# of \textsf{Google Play Store} Links & -0.81139 & 0.398 &  \\
        \# of Packages                & -0.35769 & 0.651 &  \\
        \# CI/CD Services             & -0.13900 & 0.898 &  \\
    \bottomrule
     \multicolumn{4}{l}{
         \begin{tabular}
                {@{}l@{}} $^+$Significance codes:  0 `***' 0.001 `**' 0.01 `*' 0.05 `.' 0.1 ` ' 1
         \end{tabular}
    }
    \end{tabular}
    \label{tab:model_results}
    \vspace{-7pt}
\end{table}

\begin{Summary}{}{}
Developers regularly change CI/CD pipelines, primarily focusing on setup and build phases, with most projects showing a trend toward simplification over time. While maintenance likely occurs bi-monthly on average, many projects update CI/CD pipelines more often, especially those using \textsf{GitHub~Actions}. Increased maintenance activity correlates significantly with active issue tracking, project size/age, and community engagement.
Prior research reported varying CI/CD change rates for general-purpose projects depending on the adopted CI/CD service, while change frequency over time was uncertain in those projects.
\end{Summary}

\vspace{-14pt}
\textbf{\subsection{RQ3: What are the common themes in CI/CD maintenance activities in Android apps?}}

\subsubsection{\textbf{Motivation.}}

Previous RQs have explored the complexity and evolution of CI/CD configurations. However, there is limited knowledge about the rationale behind maintenance activities developers undertake for CI/CD configurations, as well as the key aspects of the build process that they aim to improve or fix with these changes. This RQ aims to address this knowledge gap by analyzing the commit messages related to changes in \yml configuration files across various projects. The goal is to understand the motivations that drive developers to modify CI/CD configurations after their initial creation.

\subsubsection{\textbf{Approach.}}
To effectively understand the rationale behind CI/CD configuration changes, we employ both manual and automated analysis of commit messages.

\vspace{4pt}
\noindent\textbf{Manual thematic analysis of commit messages.}
We employ open coding~\cite{corbin1990grounded} to perform a thorough manual analysis of a statistically significant and stratified random sample of $384$ commit messages, based on a $95\%$ confidence level and a $\pm5$ confidence interval. Our goal in this manual analysis is to identify recurring themes and categories that explain the reasons behind CI/CD configuration changes.
Two coders (co-authors of this paper with practical experience in software engineering and CI/CD practices) participate in the manual thematic analysis of the commit messages. They engage in collaborative discussion sessions to ensure consistent interpretation and categorization. Disagreements or ambiguities in coding are resolved through consensus, with the third co-author acting as a mediator to make final decisions. This process ensures the robustness, reliability, and reproducibility of our thematic coding results.

\vspace{4pt}
\noindent\textbf{Automated analysis of commit messages.}
To further validate the results of our manual analysis, we adopt Latent Dirichlet Allocation (LDA)~\cite{blei2003latent} as a supplementary step to verify the themes identified manually. Recognizing the importance of parameter tuning in LDA for obtaining meaningful and reliable topic distributions, we follow the recommendations provided by \citet{agrawal2018wrong}. Specifically, we employ LDA with Differential Evolution (LDADE) for parameter optimization, focusing on determining the optimal number of topics and tuning the alpha and beta parameters accordingly. As a result, we obtain alpha and beta values of $0.97$ and $0.73$, respectively, and identify $10$ as the optimal number of topics.
Moreover, before performing LDA, we preprocess commit messages to ensure their suitability for topic modeling. First, URLs, email addresses, and default \textsf{GitHub} commit messages are removed. Then, stop words, numbers, and common abbreviations are filtered out, focusing on meaningful content. Lemmatization is also performed for each commit to reduce words to their root forms, and non-English characters are excluded. Also, very short messages (i.e., having fewer than three characters) are discarded as they are often non-informative and do not provide enough context. This thorough preprocessing ensures the text data is clean, standardized, and ready for effective topic modeling with LDA.
Finally, we utilize the optimized parameters and processed data to perform LDA topic modeling. For each identified topic, we extract the eight main associated keywords, assign each commit message to the topic with the highest probability of occurrence, calculate the percentage of commit messages that belong to each topic, and assess their relevance to the manually identified themes.

\subsubsection{\textbf{Findings.}}

\noindent Table~\ref{tab:themes} lists the common themes of the commit messages identified by our open coding analysis. We describe our identified themes in the following. Our analysis of open codes derived from our manual analysis of commit messages in CI/CD configurations of Android apps reveals several key insights into the current practices and trends in this area.

\begin{table}
\renewcommand{\arraystretch}{1.45}
\centering
\caption{Common themes of commit messages}
\label{tab:themes}
\begin{tabular}{L{2.5cm}  r L{5.5cm} L{4.4cm} }
\toprule
\textbf{Change Theme} & \textbf{\# (\%)} & \textbf{Description} & \textbf{E.g., Commit Message}\\
\midrule
Pipeline/Workflow Improvement  & 86 (22.8\%) & Refers to optimizations in the CI/CD pipeline workflow, such as resource usage, aiming to improve efficiency, speed, or reliability of the build process & "Use array for check workflow conditions" \\

Build Fixing	               & 57 (14.9\%) & Pertains to resolving issues that prevent successful builds, including fixing errors or bugs
in the build process or configuration & "Fixed CI / Fixed exponentially increasing broken blocks" \\

Environment Update	           & 56 (14.7\%) & Involves updates or changes to the build environment in the CI/CD process,
such as updating language versions, operating systems, or environment variables & "Add Kotlin support" \\

Testing Improvement	           & 50 (12.8\%) & Focuses on improving or enhancing the testing phase of the CI/CD pipeline,
which could include adding new tests, improving existing tests, or enhancing test automation & "Remove cache and emulator creation for testing" \\

Tool Update	                   & 48 (12.6\%) & Entails updating or changing the tools and utilities used in the CI/CD pipeline, which may
include compilers, testing tools, or deployment utilities & "Update android gradle version to alpha09" \\

Dependency Update	           & 36 (9.4\%)  & Involves updating or managing dependencies of the software project, which may
include libraries, packages, or modules that the project relies on & "Bump ruby/setup-ruby from 1.88.0 to 1.115.3 ... dependabot[bot]" \\

Deployment Improvement	       & 32 (8.4\%)  & Concerns improvements in the deployment phase of the CI/CD pipeline,
aiming to enhance deployment strategies, speed, or reliability & "Use upload-google-play@v1.0.15 to upload to Play Store" \\

Formatting Improvement	       & 13 (3.4\%)  & Refers to changes related to cleaning and organizing CI/CD configuration
file, such as indentation & "Cleanup and naming" \\

Reporting Improvement	       & 10 (2.6\%)  & Focuses on enhancing the reporting mechanisms in the CI/CD process, such
as improving feedback loops, logging, or analytics & "Fix Slack notifications" \\

Documentation Improvement	   & 4	(1.0\%)  & Involves updating the project documentation as part of the CI/CD
process, ensuring accurate and up-to-date documentation is maintained & "Update changelog action configuration" \\

License Update	               & 4	(1.0\%)  & Refers to changes made to the licensing information or files in the project, ensuring
compliance with licensing requirements & "Add Android preview license for ConstraintLayout" \\
\bottomrule
\end{tabular}
\end{table}

\vspace{4pt}
\noindent\textbf{Observation 3.1:}
\textbf{Over a third of configuration changes are concerned with improving CI workflow and fixing build issues.}
Our analysis reveals a high ratio of configuration changes allocated to \textit{Workflow Improvement} and \textit{Build Fixing} themes, accounting for 22.8\% and 14.9\% of the themes, respectively (37.4\% combined). This highlights a continual effort to refine and optimize the CI/CD process while resolving build issues, which is essential for maintaining efficient and reliable software development cycles. It also emphasizes the need for quick and frequent resolution of build issues to prevent delays in the development and deployment cycles of mobile apps, thus ensuring a smooth and continuous CI/CD process.
However, despite the invested research on build optimization and failure prediction, we observe that most of these changes are mainly performed manually, indicating a lack of tool support to improve the build process automatically. As a result, we find changes in which developers make several trials of configuration changes to assess their impact on the build workflow. For example, one of the commit messages states "\href{https://github.com/eliotstocker/Light-Controller/commit/c9a0135e33f127a3faa5b8b288f4d16af202c6cf}{\textit{attempt to give travis a little more memory by stopping some services}}" and another commit message states "\href{https://github.com/hrydgard/ppsspp/commit/c8204a24ec1fbbe750044107a24b246d7a431f4d}{\textit{Try to workaround git tag fetch failure}}".
To overcome these challenges, developers are encouraged to adopt advanced techniques to improve the monitoring and early detection of CI build issues, including those related to performance or failures. This proactive approach not only mitigates the delays in development and deployment cycles but also ensures a smoother and more reliable CI/CD process. Furthermore, this creates opportunities for researchers to evaluate the real-world applicability and efficiency of the techniques proposed in scientific research, thereby shaping future innovations in software development and boosting overall productivity.

\vspace{4pt}
\noindent
    \textbf{\textit{Compare \& Contrast with Prior Research Findings (3.1):}}
    A prior study highlighted the challenges developers face in automating CI processes and debugging build failures~\cite{zampetti2020empirical}, while another study showed that configuring CI builds to prevent failures is often related to longer build durations~\cite{ghaleb2022studying}, reflecting the effort invested in ensuring successful builds. \citet{zampetti2021ci} identified several reasons related to changes in \textsf{Travis-CI} configurations. Most importantly, they found that changes to workflow phases and sub-workflows are generally dominant, which aligns with our observation. However, other changes related to documentation and notifications were reported among frequent changes, which is inconsistent with our observation regarding these types of changes. We believe that the themes we identify capture the unique aspects of developing and maintaining Android apps, such as fixing build issues related to dependency on Android components, which differ in release requirements from other applications.

\vspace{4pt}
\noindent\textbf{Observation 3.2:}
\textbf{Over a third of configuration changes are focused on updating the build environment, tools, and dependencies.}
Our results show that CI build configurations are more frequently changed for the sake of \textit{Environment Update} (14.7\%), \textit{Tool Update} (12.6\%), and \textit{Dependency Update} (9.4\%).
Environment updates are particularly crucial as they demonstrate the necessity of maintaining current environments to align with technological advancements and security protocols. By consistently incorporating these updates, mobile apps remain compatible with the latest technologies, mitigating vulnerabilities inherent in older systems. 
Moreover, tool and dependency updates highlight the essential need for ongoing management and upgrades of CI/CD pipelines. Regular updates ensure efficient development processes, while managed dependencies keep mobile app components functional and compatible, leading to fewer interruptions and more reliable software performance.
While developers should stay updated on CI/CD trends to keep their mobile apps robust, CI/CD service providers should also ensure dependencies are secure~\cite{gruhn2013security} and suggest necessary updates.

Unlike workflow improvement and build fixing, we observe that updates to dependencies and build environment are sometimes achieved using automated bots, such as Dependabot\footnote{Dependabot: \url{https://github.com/dependabot}} and Renovate Bot\footnote{Renovate Bot: \url{https://github.com/renovatebot}}. Still, we find other commits in which such updates are performed manually, but our analysis shows that those changes do not address version upgrades but rather the way dependencies are installed or used by the CI build.
This encourages researchers to explore more on the automation of dependency management and upgrades in the context of CI/CD to focus more on practical aspects that benefit both the open-source and industry communities.
Overall, this finding highlights the rapid advancement of technology, emphasizing the importance of staying up-to-date as a key strategy for minimizing risks, ensuring compatibility, and maintaining a competitive edge in the industry.

\vspace{4pt}
\noindent
    \textbf{\textit{Compare \& Contrast with Prior Research Findings (3.2):}}
    Prior studies~\cite{zampetti2021ci,widder2019conceptual,khatami2024catching,gallaba2018use} identified various patterns of changes related to the restructuring of CI configurations. These patterns address multiple aspects like performance, security, and organization, which are often linked, either directly or indirectly, to CI environments and dependencies. Our analysis, however, highlights that changes to the environment and dependencies are particularly prevalent in Android apps, suggesting rapid, ongoing evolutions in the platforms and libraries that support these types of applications.

\vspace{4pt}
\noindent\textbf{Observation 3.3:}
\textbf{\textit{Testing Improvement} and \textit{Deployment Improvement} changes account for 12.8\% and 8.4\%, respectively.}
Robust testing and regression in CI/CD pipelines ensure early detection and resolution of bugs, leading to the development of more reliable and high-quality mobile apps. Similarly, deployment ensures rapid and reliable and efficient release cycles, thus reducing the time-to-market for mobile app updates.
Despite the importance of frequently maintaining testing and deployment, our results show relatively fewer changes to them compared to other practices, which could be attributed to several factors.
These processes might already be well-established and mature within the organization, hence requiring fewer changes and improvements over time, such as enable more features (e.g., \href{https://github.com/cuongpm/android-minimal-todo/commit/3dc9cee50ba6c5bd310d6bbfcce3867d0254cc20}{\textit{add test coverage}} and \href{https://github.com/niccokunzmann/mundraub-android/commit/2343d9357844f49ad55122b1579e4474f941a061}{add deployment to playstore}) or fixing minor relevant issues (e.g. \href{https://github.com/compscidr/hello-java-android/commit/b60480cb7d4ea7bb2a314ebf60b859b07863c6c4}{try number two for multiple devices} and \href{https://github.com/infinitepower18/WSA-InstalledApps/commit/77d6da33a0f2a4750c88f982bd1d1b838fac5820}{fix artifact name}).
In some cases, testing and deployment might not be considered immediate strategic priorities for developers compared to other areas such as feature development, user experience, or performance optimization (e.g., \href{https://github.com/jpaulynice/android-jigsaw-puzzle/commit/91660f71970bf4c0aa47440b6af1809f880edfed}{committing to fix tests later}).
Still, developers are advised to continuously monitor testing and deployment processes, given the critical importance of robust testing and deployment in mobile app deployment, making ongoing improvements and updates are essential to adapt to new technologies and changing requirements.
Likewise, researchers should pay more attention to developing techniques that recommend potential co-changes in testing or deployment configuration in response to changes in source code.

\vspace{4pt}
\noindent
    \textbf{\textit{Compare \& Contrast with Prior Research Findings (3.3):}}
    Prior research highlighted the crucial role of testing in CI and identified several configuration patterns related to defining, executing, improving reliability, and optimizing automated test performance in the CI pipeline. However, there was no direct link established between testing and its corresponding changes in CI configurations. Yet, given that developers may use external scripts to run CI testing~\cite{zampetti2021ci,gallaba2018use}, we expect that changes to testing might also be external. Nevertheless, our results still reveal some ratio of changes to testing, though relatively low. For deployment, contrary to our observations, prior research indicated that changes in deployment are somewhat frequent~\cite{gallaba2018use}, which does not seem to apply to Android apps, likely due to the preference for manual deployment.

\vspace{4pt}
\noindent\textbf{Theme Validation via LDA.}
To further support the relevance and accuracy of themes identified in our manual analysis across all commits in our dataset, we perform LDA to generate topics automatically.
The automatically generated topics by LDA align with our manually identified themes and provide broader insights into diverse CI/CD maintenance activities.
Table~\ref{tab:LAD_topics} shows the $10$ topics generated by LDA, each with the most common keywords, an example commit message, and a brief description of each topic. We can see that the topics generated by LDA are consistent with the themes of the commit messages that we identify manually. 
For instance, topics $\#1$ and $\#5$ (with 25.5\% of the commits) align with the \textit{Workflow/Pipeline Improvement} and \textit{Build Fixing} themes, despite the difference in proportion.
Topics $\#5$, $\#6$, $\#7$, and $\#8$ (with 36.8\% of the commits) correspond closely with the \textit{Environment Update}, \textit{Tool Update}, and \textit{Dependency Update} themes, indicating a strong emphasis on keeping the build process up-to-date. We observe the reliance on the \textit{Renovate} (9.2\%) and \textit{Dependabot} (6.9\%) bots for automated dependency updates, while many other updates could potentially be performed manually. This makes it important for developers to be more aware of the availability and benefits of automated tools for various CI/CD improvements, and also signals to researchers the need to address the existing gaps in this context.
With respect to testing and deployment, we observe topics $\#2$, $\#3$, $\#9$, and $\#10$ (with 37.7\% of the commits) addressing these processes, with more focus toward improving testing (three topics with 31.5\%) over deployment (only 6.2\%).
We further emphasize the critical role of deployment in mobile apps and recommend future research to explore the correlation between the frequency of deployment maintenance and the quality of delivered apps and user satisfaction, to determine whether it leads to enhancement or deterioration.

\begin{table}[ht]
    \renewcommand{\arraystretch}{1.5}
    \centering
    \vspace{-5pt}
    \caption{LDA generated topics representing all commit messages in our dataset}
    \vspace{-5pt}
    \label{tab:LAD_topics}
    \begin{tabular}{c L{5.5cm} L{6.3cm} r}
    \toprule
     & \textbf{Keywords} & \textbf{Description} & \textbf{\# (\%)}\\
    \midrule
    1  & \texttt{\textbf{fix}, updated, version, flutter, \textbf{error}, \textbf{issue}, minor} & Fixes errors and issues & 4,195 (16.1\%) \\

    2  & \texttt{\textbf{test}, added, mobile, local, report, api} & Add testing features & 3,272 (12.6\%) \\
    
    3  & \texttt{\textbf{build}, tool, \textbf{run}, \textbf{android test}, configure, \textbf{fix}, jdk} & Testing tool and execution maintenance & 3,242 (12.4\%) \\
    
    4  & \texttt{android, changed, \textbf{install}, sdk, change, \textbf{dependency}, added} & Dependency and environment updates & 3,154 (12.1\%) \\
    
    5  & \texttt{\textbf{workflow}, \textbf{cache}, config, make, \textbf{pipeline}, step, refactor} & Workflow and pipeline enhancements & 2,448 (9.4\%) \\
    
    6  & \texttt{action, chore deps, co-authored-by, \textbf{renovatebot}, v3, v2, checkout} & Automated dependency updates by Renovate Bot & 2,397 (9.2\%) \\
    
    7  & \texttt{java, version, gradle, \textbf{upgrade}, use, node, set} & Environment and tool upgrades & 2,247 (8.6\%) \\
    
    8  & \texttt{\textbf{release}, note, signed-off-by, \textbf{dependabot bot}, \textbf{auto-update-dependencies}, release\_note, changelog} & Automated dependency updates and documentation by Dependabot & 1,792 (6.9\%) \\
    
    9  & \texttt{name, try, \textbf{integration}, \textbf{android test}, rename, fix, \textbf{coverage}} & Testing-related fixing & 1,683 (6.5\%) \\
    
    10 & \texttt{\textbf{upload}, github, request, merge, setup, \textbf{apk}, change-id} & Packaging and deployment updates & 1,625 (6.2\%) \\
    \bottomrule
    \end{tabular}
\end{table}

\vspace{10pt}
\begin{Summary}{}{}
Over a third of CI/CD configuration changes focus on improving workflow and fixing build issues, while another third target updates to the build environment, tools, and dependencies. The automatically generated topics align with our manually identified themes, offering broader insights into the diverse maintenance activities within CI/CD processes.
This is consistent with some prior findings on general-purpose projects that highlight workflow restructuring and environment updates as common CI change motivations, but still inconsistent with other findings that report more frequent changes related to documentation, notifications, and deployment than we observed in Android projects.
\end{Summary}

\vspace{4pt}
\section{Implications} \label{sec:Implications}
This section discusses the implications of our findings for developers, researchers, and CI/CD service providers.

\vspace{4pt}
\noindent\textbf{The need for standardized practices in CI/CD configurations.}
Our results in \textit{Observation 1.1} show a lack of standardized directives and values, indicating a need for uniform practices in configuration file conventions. This highlights the varied usage of CI/CD configuration practices, thus underlining the need for standardization. Though Java and Kotlin are found to share common directives across CI/CD services, this is not the case with \textsf{Travis~CI} reflects service-specific requirements or optimizations for these languages. This suggests that Java and Kotlin projects can leverage shared best practices, tools, or scripts. Therefore, for any CI/CD service, developers may need to tailor configurations to specific service requirements, which also alarms the need for researchers to develop more automated approaches for that purpose.
Building on this, we argue that standardized practices---such as improved Android-specific documentation, curated configuration templates, reusable components like \textsf{GitHub Actions} workflow templates\footnote{\url{https://docs.github.com/en/actions/writing-workflows/using-workflow-templates}} and \textsf{CircleCI} orbs\footnote{\url{https://circleci.com/orbs}}, and Android-focused best-practice guidelines---can help address the lack of consistency observed across services. These practices can collectively contribute to reducing configuration effort, minimizing errors, improving maintainability, ensuring consistency, and simplifying migration across CI/CD tools, ultimately lowering the barrier to effective CI/CD adoption in Android development. Moreover, new developers or team members may face difficulties in understanding and contributing to projects with varying CI/CD configurations, leading to longer onboarding times and challenges in knowledge transfer. These factors can negatively affect team efficiency and productivity, further reinforcing the need for standardized CI/CD practices, especially in the context of Android app development.

\vspace{4pt}
\noindent\textbf{Incorporating CI/CD pipeline optimization mechanisms in practice.} To support developers in adopting CI/CD practices, release management tools should offer automated solutions for configuring and optimizing CI/CD pipelines.
In \textit{Observation 1.3}, we find that projects use various app testing practices, including different testing tools and platforms, and the selection of appropriate emulators for their apps. However, traditional software testing and emulator use often pose delays to CI/CD pipelines. Despite extensive research on the selection and prioritization of devices for app testing~\cite{DBLP:conf/sigsoft/KhalidNSH14,DBLP:journals/sigsoft/Yang24} or test cases~\cite{DBLP:conf/sigsoft/ElbaumRP14,DBLP:journals/emse/Pan2022}, these approaches are not being applied in our studied projects.
Successful adoption of advanced and recent testing mechanisms can help provide timely feedback to app developers about the quality (e.g., test coverage) of new app releases. Future research should focus more on developing automated testing frameworks that can easily align and integrate with existing CI/CD services and testing platforms to increase their chances of being adopted in practice. This may also include tools that automatically generate test cases or integrate test changes into the CI/CD pipeline.

\vspace{4pt}
\noindent\textbf{Tailored CI/CD maintenance strategies.}
Our findings in \textit{Observation 2.1} and \textit{Observation 2.2} reveal substantial variation in the frequency and nature of CI/CD configuration updates, highlighting the need for tailored maintenance strategies. While some projects exhibit regular, incremental updates, others show infrequent or cumulative changes. This suggests that one-size-fits-all approaches to CI/CD maintenance are ineffective, and tools should be adaptable to the unique needs of projects.
Factors such as project complexity, issues, popularity, and historical activities can significantly influence the rate and nature of CI/CD modifications, as observed in \textit{Observation 2.5}. Moreover, the possibility of new developers joining projects necessitates a flexible approach to maintenance, as it can introduce overhead to CI/CD update practices. Hence, both researchers and CI/CD service providers are encouraged to consider developing customizable maintenance mechanisms that rely on historical events of software projects, to align with the varied practices, ensuring CI/CD configurations remain efficient and responsive to project demands.

\vspace{4pt}
\noindent\textbf{Recommendation mechanisms for updating CI/CD configurations.} It is important to emphasize the critical role of automation tools, especially those that utilize bots, in maintaining up-to-date CI/CD pipelines. Our results in \textit{Observation 3.2} suggest an extensive use of bots, such as Dependabot, to automatically update software dependencies. It is also necessary to explore how these mechanisms can address broader CI/CD configuration updates, beyond mere dependency updates. Version control systems, such as \textsf{GitHub}, alongside CI/CD services, should offer intelligent recommendation systems to autonomously propose configuration changes, aligning with the latest best practices and standards to ensure effective and efficient workflows. As \textit{Observation 3.1} indicates, developers often undertake numerous changes to improve the CI workflow and resolve build issues, which can span many stages of CI/CD pipelines.
This gives an opportunity for advanced automated tools, leveraging generative AI and large language models, to systematically identify, implement, and refine improvements.
Therefore, an investment in such intelligent tools is not only a pathway to minimizing manual interventions but also a strategy for keeping pace with innovative approaches in the ever-evolving realm of software development and integration.
Researchers should also be part of this effort and should pay considerable attention to developing sophisticated techniques that align with the practical needs of software developers and build engineers.

\vspace{4pt}
\noindent\textbf{More attention in testing and deployment phases of CI/CD pipelines.}
As \textit{Observation 2.4} demonstrates, while CI/CD pipeline maintenance activities primarily address setup and build processes, testing and deployment remain less frequently modified, indicating limited continuous improvement in these critical phases. This gap may be partly due to manual intervention or a lack of robust automation tools.
The challenge is even greater in Android apps, where testing often depends on emulators to support a wide range of devices and OS versions. These requirements complicate integration with third-party services, such as \textsf{Google Play Console}\footnote{\url{https://play.google.com/console}}and \textsf{AWS Device Farm}\footnote{\url{https://aws.amazon.com/device-farm}}, creating further barriers to automation.
Deployment adds another layer of complexity, as Android apps must be pushed to app stores like the \textsf{Google Play Store}, where updates undergo review processes involving security, compatibility, and privacy checks, unlike web applications, which can be updated directly. Given these constraints, CI/CD services such as \textsf{GitHub~Actions} often recommend performing app deployment manually and externally from the main CI/CD workflow.
Therefore, future research and CI/CD service providers should explore advanced automated solutions, such as AI-driven test generation~\cite{wang2024software,su2024enhancing,schafer2023empirical} and optimized deployment strategies~\cite{vadde2022ai}, to reduce manual overhead, improve feedback cycles, and enhance the overall effectiveness of mobile CI/CD pipelines. Promising directions include automated test case generation tailored to CI/CD pipelines beyond prompt engineering that relies solely on the latest error messages, but is rather based on log histories and previous fixes~\cite{tufano2019learning}. Example tools include \textsf{Launchable}\footnote{\url{https://www.launchableinc.com}} and \textsf{Diffblue}\footnote{\url{https://www.diffblue.com}}, neither of which is adopted in our dataset.
In terms of deployment, to enhance existing support by platforms like \textsf{GitHub} in generating release notes\footnote{\url{https://docs.github.com/en/repositories/releasing-projects-on-github/automatically-generated-release-notes}}, AI could further help in release readiness assessment~\cite{hoang2020cc2vec} and detection of compliance risks by analyzing code changes and issue history~\cite{yang2025code}. These capabilities could be integrated into CI/CD platforms as reusable workflows to allow adaptation to project-specific configurations over time. Despite the existence of paid AI-driven tools for software deployment, such as \textsf{Digital.ai}\footnote{\url{https://digital.ai}}, and research-based tools, such as \textsf{LADs}~\cite{faraz2025lads}, our dataset does not show any utilization of such tools.
By integrating AI in existing tools and research, these directions offer practical and impactful ways to reduce manual effort, enhance reliability, and standardize CI/CD practices across Android projects.
Furthermore, open-sourcing such tools could further benefit the software engineering community, particularly for teams with limited budgets or expertise.

\section{Threats to validity} \label{sec:Threats}

\subsection{Construct Validity}
Construct threats to validity are concerned with the degree to which our analyses measure what we intend to analyze.
In our study, we rely on data collected mostly from \textsf{GitHub} and analyzed using Python scripts written by the authors of this paper. Mistakenly computed values may affect our conclusions. To address this concern, we carefully filter and verify the data to minimize the possibility of computational errors that might impact our analyses.
In addition, we use open coding to manually categorize the configuration directives into their respective phases or roles, which could be subjective. To mitigate this threat, multiple iterations of manual and automated analyses are performed by two co-authors of this paper with practical experience in software engineering and CI/CD practices to validate the generated categories using statistically significant random samples of CI/CD configurations.
This process of validation, refinement, and re-evaluation was repeated iteratively six times for the remaining unfiltered projects until we reached an observed agreement of $93\%$ in manually validating the automatically filtered projects, ensuring that our final dataset reliably excluded toy projects and retained real Android mobile apps.
Similarly, we perform a thematic analysis of commit messages to understand what maintenance activities developers undertake on CI/CD pipelines. Given that we are not the owners of the analyzed apps, there may be a misunderstanding of changes made in the studied commits. To address this, the analysis is also validated by two co-authors and further verified using automated topic modeling via LDA.
Lastly, the reported results and statistics are calculated using our developed Python scripts. Though these scripts may contain defects that could lead to improper interpretations, we verify them by cross-checking the results with samples from our dataset.

\subsection{Internal validity:}
Internal threats to validity are concerned with the ability to draw conclusions from our empirical results.
In our study, we investigate the association between project characteristics and maintenance activity in terms of CI/CD commit frequency using OLS linear regression. We recognize that our current variables may not fully capture all relevant factors, thus incorporating additional variables might change our conclusions and provide deeper insights, which we recommend for future research.
In addition, we use LDA topic modeling for automatically identifying patterns in CI/CD commit messages. However, this could lead to generating incoherent or misleading topics due to poor parameter choices. To address these, we perform data preprocessing, tune LDA parameters (such as the number of topics, alpha, and beta), and manually refine the topic distributions. This ensures that our obtained results are meaningful, reliable, and aligned with our aim of validating manually identified themes of commit messages.

\subsection{External validity:}
External threats are concerned with our ability to generalize our study results.
In our study, we initially collect 6,322 open-source Android apps that adopt CI/CD, relying on two sources that are commonly used in prior research: \textsf{GitHub} and \textsf{F-Droid}.
When determining whether an app adopts CI/CD, we focus on well-known CI/CD services (e.g, \textsf{GitHub~Actions}, \textsf{Travis~CI}, \textsf{CircleCI}, and \textsf{GitLab~CI/CD}) and their associated \yml configuration files. We excluded CI/CD services for which we find only a limited number of adopting apps. However, some of the excluded apps may still use other CI/CD services that we did not account for. Also, some apps might not contain \yml files, yet adopt CI/CD with default configurations or configured through web interfaces. Hence, our dataset might overlook some relevant apps, but those can be challenging to identify.
Moreover, we realize that a considerable number of apps initially included in our dataset, which satisfied our filtration criteria, are not real apps but rather examples or samples provided as part of libraries or for educational purposes. We refer to these as toy projects. Therefore, to mitigate any possible threat to these apps, we perform an additional filtration step to exclude these types of apps from our dataset through rigorous analysis, resulting in a final set of 2,557 apps.
Furthermore, our dataset is restricted to open-source Android apps, excluding non-free, iOS, and non-Android apps, which limits the generalizability of our results. A major challenge in analyzing non-free apps is their limited public availability.
In addition, the lack of comparisons with non-Android projects prevents us from drawing broader conclusions about CI/CD adoption, configuration complexity, and maintenance practices across domains. To partially address this threat, we compare and contrast our findings with prior research findings on general-purpose CI/CD practices, allowing us to situate our observations within broader trends. Nevertheless, our findings may still benefit developers using various CI/CD services, as common challenges, such as dependency management and testing, are widely encountered. Therefore, we aim to further expand our analyses and encourage future research to include a broader range of application domains to support more generalizable conclusions and to explore domain-specific patterns in CI/CD configuration and maintenance practices.

Our study focuses exclusively on Android apps, which may limit the generalizability of our conclusions across other application domains. While this focus allows for a deep investigation of mobile-specific CI/CD practices, we acknowledge that comparisons with non-Android projects, particularly regarding CI/CD adoption, configuration complexity, and maintenance frequency, could strengthen our findings. We aim to expand our analyses in future work to include a broader range of application types to support more generalizable conclusions.

\section{Related work} \label{sec:RelatedWork}
This section presents prior work related to our study, including those studying the practices in mobile app releasing and CI/CD adoption.

\subsection{Studies on analyzing mobile app releases}

Several studies investigated the characteristics of bad releases and emergency releases of popular apps in the \textsf{Google Play Store}~\cite{Bad_Updates_TSE,Emergency_Updates_EMSE,Islem_EMSE}.
They found that non-functional issues (e.g., UI issues) can lead to bad/emergency releases. 
\citet{DBLP:journals/ese/McIlroyAH16} studied the relationship between the release cycles in mobile apps and the app ratings.
They found that frequently releasing apps have higher ratings than apps with slower release cycles. 
\citet{DBLP:conf/sigsoft/Dominguez-Alvarez19} found that Android apps have more releases than \textsf{iOS} apps. 
\citet{DBLP:journals/ese/YangHZH22} studied the release notes of popular apps in the \textsf{Google Play Store} and also surveyed 102 app developers. They found that, while participants consider release notes as a useful artifact to inform users about changes, there are variations in the practices that developers follow to create and use these notes.

\citet{DBLP:conf/wcre/NayebiAR16} studied the release practices for mobile apps by surveying app developers and app users. 
They found that 53\% of the participant developers follow a specific release practice (e.g., deploy their releases weekly). 
They also found that release practices have a potential impact on the success of an app.  
Later, \citet{DBLP:conf/esem/NayebiFR17} studied 11,514 releases of 917 open-source applications, and found that 44.30\% of releases are not deployed to the Google App Store. 
They built models to predict whether the generated release will be deployed to the app store.

Unlike previous studies, this paper analyzes the challenges in CI/CD configurations that could potentially influence the above releasing issues. We explore how these configurations still rely on manual or external releasing mechanisms, which might contribute to deployment delays. We provide insights into optimizing release practices using automated techniques for better app store delivery.

\subsection{Studies on CI/CD adoption and configurations}
Martin Fowler proposed ten best practices for adopting CI in software projects~\cite{Fowler_CI}, which researchers then studied to understand the impact of these practices in real-world projects~\cite{DBLP:conf/kbse/ZhaoSZFV17}. 
\citet{DBLP:conf/kbse/HiltonTHMD16} analyzed thousands of \textsf{GitHub} projects and surveyed 442 software developers on CI adoption challenges, and discovered that only 40\% of the projects use CI, with the primary barrier being the lack of experience of developers.
\citet{widder2019conceptual} also identified CI adoption challenges by reviewing 37 studies and surveying 132 developers, and reported common difficulties in troubleshooting, inconsistency of tools across CI/CD services, and complex setups.
\citet{DBLP:journals/corr/abs-2103-04251} also studied the challenges in CI adoption, and found that automating UI testing and pull request review bottlenecks represent the main issue.
\citet{zampetti2020empirical} developed a catalogue of bad CI/CD practices from developer interviews and Stack Overflow analysis, highlighting mismanaged branching and versioning, and unaddressed infrastructure, configuration, and testing smells.
\citet{vassallo2020configuration} studied CD configuration issues and developed a tool to detect relevant configuration smells.

\citet{bouzenia2024resource} examined resource usage in \textsf{GitHub~Actions} and found CI/CD pipelines resource-intensive, particularly in testing and building.
\citet{jin2021helped} assessed CI improvement techniques and reported trade-offs in cost, fault detection, and feedback time.
\citet{ghaleb2019empirical,ghaleb2019studying,ghaleb2022studying} identified common patterns and trade-offs in configuring CI pipelines, specifically concerning their correlations with build durations and failures.

\citet{rzig2022characterizing} also investigated CI adoption issues, finding lower CI adoption rates in ML projects compared to other software. They identified that common CI challenges, like debugging build failures, are worsened by ML's specific needs, such as large datasets.

Unlike previous studies, this paper investigates the complexity and challenges of configuring CI/CD pipelines in open-source Android apps. We perform qualitative and quantitative analyses and provide insights into the common practices and the range of maintenance activities associated with CI/CD in this area. We also identify inconsistencies and a lack of use of sophisticated testing techniques and deployment mechanisms as part of CI/CD processes.

\subsection{Studies on CI/CD for mobile apps}
There is limited research attention related to the adoption of CD practices in mobile apps. 
For example,~\citet{DBLP:journals/smr/LiuLLMG24} studied the evolution of build systems in 5,222 Android apps. 
They found that Gradle is the most adopted build system among the studied apps, and also observed that only 31\% of the projects using the Gradle build system can be built successfully.
Similarly, \citet{DBLP:conf/mobilesoft/PoleseHT22} found that 31.4\% of the studied apps contain versions of source code that fail to build, and identified five primary reasons for these build failures.

\citet{DBLP:conf/icse/KlepperKPBA15} proposed the Rugby process as an agile process with support for continuous delivery. 
They applied the Rugby process to the different case studies of mobile apps. 
\citet{DBLP:conf/sigsoft/RossiSSBSS16} demonstrate their experience of adopting the testing and continuous deployment practices in Facebook.
Later,~\citet{DBLP:journals/ese/CruzAL19} studied testing practices in open-source apps.
They found that continuous integration and continuous deployment pipelines are not widely adopted in the studied mobile apps.
Recently,~\citet{DBLP:conf/esem/WangZXY23} studied the adoption of UI testing in 318 \textsf{Google Play Store} apps deployed on \textsf{GitHub}. 
Though they observed that 40\% of the studied apps use CI/CD services (e.g., \textsf{GitHub~Actions} and \textsf{Travis~CI}), they found that the adoption of CI/CD services is primarily used for unit testing.
Similar to our study, \citet{DBLP:conf/kbse/LiuSZL0022} analyzed 7,899 Android apps hosted on \textsf{GitHub}, \textsf{Bitbucket}, and \textsf{GitLab} that adopt CI/CD services, with \textsf{GitHub~Actions} and \textsf{Travis~CI} as the most commonly adopted. 
However, their work mainly offers general statistics about the adoption of CI/CD services without delving deeper into specific configurations or maintenance practices across CI/CD services. Furthermore, their dataset was not rigorously filtered to ensure that only actual Android apps were included.

This paper complements previous research by conducting more in-depth analyses of CI/CD pipeline configuration practices and maintenance activities. In contrast to the above studies, our investigation spans across four different CI/CD services, namely \textsf{GitHub~Actions}, \textsf{Travis~CI}, \textsf{CircleCI}, and \textsf{GitLab~CI/CD}, on a carefully curated set of $2.5k$ Android apps collected from \textsf{GitHub} and \textsf{F-Droid}. This enabled us to uncover the absence of commonalities among these services and identify potential trade-offs when selecting a CI/CD service for a specific project.

\section{Conclusion} \label{sec:Conclusion}
This paper provides a comprehensive analysis of CI/CD practices in Android apps, revealing the complexities and challenges developers face due to a lack of standardization in CI/CD configurations across the available services. Our empirical results suggest a significant focus on the setup phase in most CI/CD pipelines, with limited emphasis on testing and deployment. The findings also indicate that active maintenance, though infrequent, correlates with factors like project maturity and community engagement. Our findings agree with prior research on general-purpose CI/CD practices regarding configuration complexity, but are somewhat inconsistent with certain aspects of configuration change frequency and themes. Overall, our study highlights the need for standardized practices to improve interoperability and suggests that automation and AI-driven solutions could significantly enhance CI/CD processes. Furthermore, it is recommended to establish adaptable frameworks and open-source tools to overcome resource constraints, particularly in testing and deployment. Addressing these aspects could facilitate smoother adoption and maintenance of CI/CD pipelines in mobile development, ultimately leading to more efficient and effective software delivery.

In the future, we aim to seek feedback from software developers on our empirical findings to guide future research into the areas of highest demand for automation in CI/CD processes. We also aim to expand our analyses in future work to include a broader range of application domains to support more generalizable conclusions and to explore domain-specific patterns in CI/CD configuration and maintenance practices. Moreover, we plan to study how deficiencies in CI/CD pipelines are linked to mobile app success, particularly user perception (e.g., app ratings in app stores).

\section*{Acknowledgment} \label{sec:Acknowledgment}
We acknowledge the support of the Natural Sciences and Engineering Research Council of Canada (NSERC), \textbf{[RGPIN-2021-03969]}.

\section{Data Availability}
\label{DataAvailability}
The replication package of our study (including the data, code, and raw results) is available on Figshare~\cite{our_replication_package}.

\bibliographystyle{ACM-Reference-Format}
\bibliography{paper}{}

\end{document}